\documentclass[12pt,preprint]{aastex}

\newcommand{\grs}      {\mbox{\rm\,GRS~1758--258}}
\newcommand{\onee}     {\mbox{\rm\,1E~1740.7--2942}}
\newcommand{\gx}       {\mbox{\rm\,GX~339--4}}
\newcommand{\cyg}      {\mbox{\rm\,Cyg~X--1}}
\newcommand{\cygth}      {\mbox{\rm\,Cyg~X--3}}
\newcommand{\mdotd}     {\dot{M}_{\rm D}}
\newcommand{\mdoth}     {\dot{M}_{\rm H}}

\begin{document}

\title{Two Different Long-term Behaviors
in Black-Hole Candidates: Evidence for Two Accretion Flows?}

\author{D. M. Smith\altaffilmark{1},
W. A. Heindl\altaffilmark{2}, J. H. Swank\altaffilmark{3}}

\altaffiltext{1}{Space Sciences Laboratory, University of California Berkeley, 
Berkeley, CA 94720}
\altaffiltext{2}{Center for Astrophysics and Space Sciences, Code 0424, University
of California San Diego, La Jolla, CA 92093}
\altaffiltext{3}{NASA Goddard Space Flight Center, Code 666, Greenbelt, 
MD 20771}

\begin{abstract}

We discuss the results of long-term hard x-ray monitoring of Galactic
black-hole candidates \onee, \grs, \cyg, \gx, and \cygth\ with the \it
Rossi X-Ray Timing Explorer (RXTE)\rm.  The objects divide into two
classes.  In the first class, exemplified by \cyg, luminosity and
spectral hardness evolve simultaneously.  In the second class, the
relation is more complicated: the softest spectra occur while the
count rate is dropping.  Most models of accretion, tailored
to \cyg, do not predict the second sort of behavior.  One 
interpretation is a simple model with two simultaneous,
independent accretion flows: a thin disk and a hot halo.  A drop in
the accretion rate affecting both flows would propagate through the
halo immediately but might take up to several weeks to propagate
through the disk.  While the inner halo is thus temporarily depleted
compared to the disk, a temporary soft state is expected.
This picture is supported by the observation that those
sources which show delays (\onee, \grs, and \gx) are expected to have
low-mass companions, and those which do not (\cyg, \cygth) are known
or thought to have high-mass companions.  Low-mass companions imply
accretion by Roche-lobe overflow, with a high specific angular
momentum in the accreting material, and therefore a large disk with a
long viscous timescale.  Wind accretion from massive companions is
expected to result in a much smaller disk, and thus little viscous
delay.

\end{abstract}


\section{Introduction}

Among the x-ray binaries which are identified as black-hole candidates
by their mass functions, only three show persistent bright emission:
\cyg, LMC~X--3, and LMC~X--1.  At least two more non-transient
Galactic objects are usually considered black-hole candidates by
courtesy due to the similarity of their hard x-ray behavior to that of
\cyg: \onee and \grs. \gx occupies a position intermediate between
persistent and transient sources, being bright a significant fraction
of the time but quiescent more often.  The optical counterparts of
the latter three systems have not been detected due to their distance,
extinction by Galactic dust, and, in the case of \gx, dominance
of the accretion-disk's optical emission.

\grs\ almost certainly has a low or intermediate mass companion, and is
therefore likely to be accreting by Roche-lobe overflow, in contrast
to \cyg, which is thought to accrete the wind from its O supergiant
companion.  K-band observations of \grs\ \citep{Ma98} show that the
two brightest stars consistent with the Very Large Array (VLA) radio
position appear to be a low-mass K-type red giant and a main-sequence
F star; more massive stars, which might fuel wind accretion, are ruled
out.

For \onee, \citet{Mi91} suggested a companion later than B2, but their
argument was indirect.  The upper limit was on 6~cm radio flux; UV
from an earlier companion would produce so much ionization in a
positionally coincident molecular cloud that a compact HII region
would have been seen with the VLA.  This argument relies on \onee\
being within the cloud and not behind or in front of it.  HCO$^+$
emission-line observations have been used to argue that it is indeed
within the cloud \citep{Ph95,Ya97}, but the small amplitude of the
6.4~keV iron fluorescence line from the source has been used to argue
that it is behind the cloud \citep{Ch96} or in front of it
\citep{Sa99}.  The deepest K-band observations near \onee\ were
recently made with the Very Large Telescope (VLT) of the European
Southern Observatory \citep{Ma00}.  Four stars of K magnitude
18.5-19.0 are about 1'' from the VLA position, and each might
represent a massive companion.  \citet{Ma00} found that the K-band
3$\sigma$ upper limit exactly at the VLA position (19.5) would
constrain the companion to be a main sequence star later than about
B8~V or a giant earlier than G5~III.  The conclusion awaits
improvement of the 1'' uncertainty in the alignment of the VLA and VLT
fields.  Since \onee\ shares the characteristic of bright, extended
radio jets with \grs\ and not \cyg, one might argue (independent of
the evidence presented below) that it is more likely to be similar to
\grs\ overall, i.e. to be a Roche-lobe accretor.

Optical emission has been seen frequently from \gx, which shows less
absorption than either \onee\ or \grs, and is probably closer.  The
optical spectrum has always been consistent with an accretion disk,
however, so there has been no spectral classification of the
companion.  Optical observations taken when the x-ray luminosity was
very low \citep{Ca92} imply an upper limit of (0.6 $\pm$ 0.2)
L$\sun$ for the companion at the farthest estimated distance of
4~kpc \citep{Co87,Zd98} and one tenth of that at the closest
estimated distance of 1.3 kpc \citep{Pr91}.  \citet{Zd98} derived a
somewhat higher extinction (A$_v$ = 3.4 - 4) than that taken by
\citet{Ca92} from \citet{Il86} (A$_v$ = 2.2), which would increase the
upper limit to about 3 L$\sun$.  Even at this highest upper limit,
the companion cannot be a high-mass star providing a significant wind.
 
The unique x-ray binary \cygth\ has been interpreted
as a black hole in a 4.8 hr orbit with a Wolf-Rayet star 
\citep{Sc96,Ke96}, but this identification of both components 
has been questioned \citep{Mi96,Mi98}.  A Wolf-Rayet star
could provide sufficient material in its wind to feed a
bright x-ray binary system.

\cyg, \onee, \grs, and \gx\ spend most of their time in the spectral
state called ``hard'' or ``low''.  This state, one of the two
canonical states of black-hole candidates at moderate luminosity, is
characterized by a hard power law (photon index 1.4 to 1.9) with an
exponential rollover around 100 keV.  The other canonical state,
``soft'' or ``high'', is characterized by a softer power law with no
exponential cutoff, plus a thermal component with a temperature on the
order of 1 keV which can dominate the overall luminosity.  There can
be a gradual transition between these canonical states, now usually
called an intermediate state \citep{Me97}.  It has recently been
discovered, however, that at least one property of this state is not
intermediate at all: \citet{Po00} found that the lags between rapid
variations in soft and hard flux are far longer during state
transitions of \cyg\ than in either the soft or hard states.  The
``very high'' state, characterized by high accretion rates and more
complicated behaviors, is beyond the scope of this paper, never having
been displayed by \cyg, \onee\ or \grs.

One common picture of accretion in these systems
is a standard thin accretion disk at large radii, which is
truncated at some inner radius and replaced by a hot, more spherical
flow \citep[e.g.,][]{Sh76}.  The thin disk produces thermal emission
and the hot flow scatters it up into a power law.  The transition from
the low state to the high state is thought to occur when the accretion
rate increases, with the result that the transition radius moves
inward until the thin disk extends all the way down to the last stable
orbit allowed by general relativity \citep[e.g.,][]{Es98,Ja00}.

In this paper we will demonstrate that this picture, although adequate
to explain the behavior of \cyg, will have to be altered to
address the more complicated evolution of luminosity and spectrum in
other black-hole candidates.

\section{Results}

\subsection{Observations}

\begin{figure}
\plotone{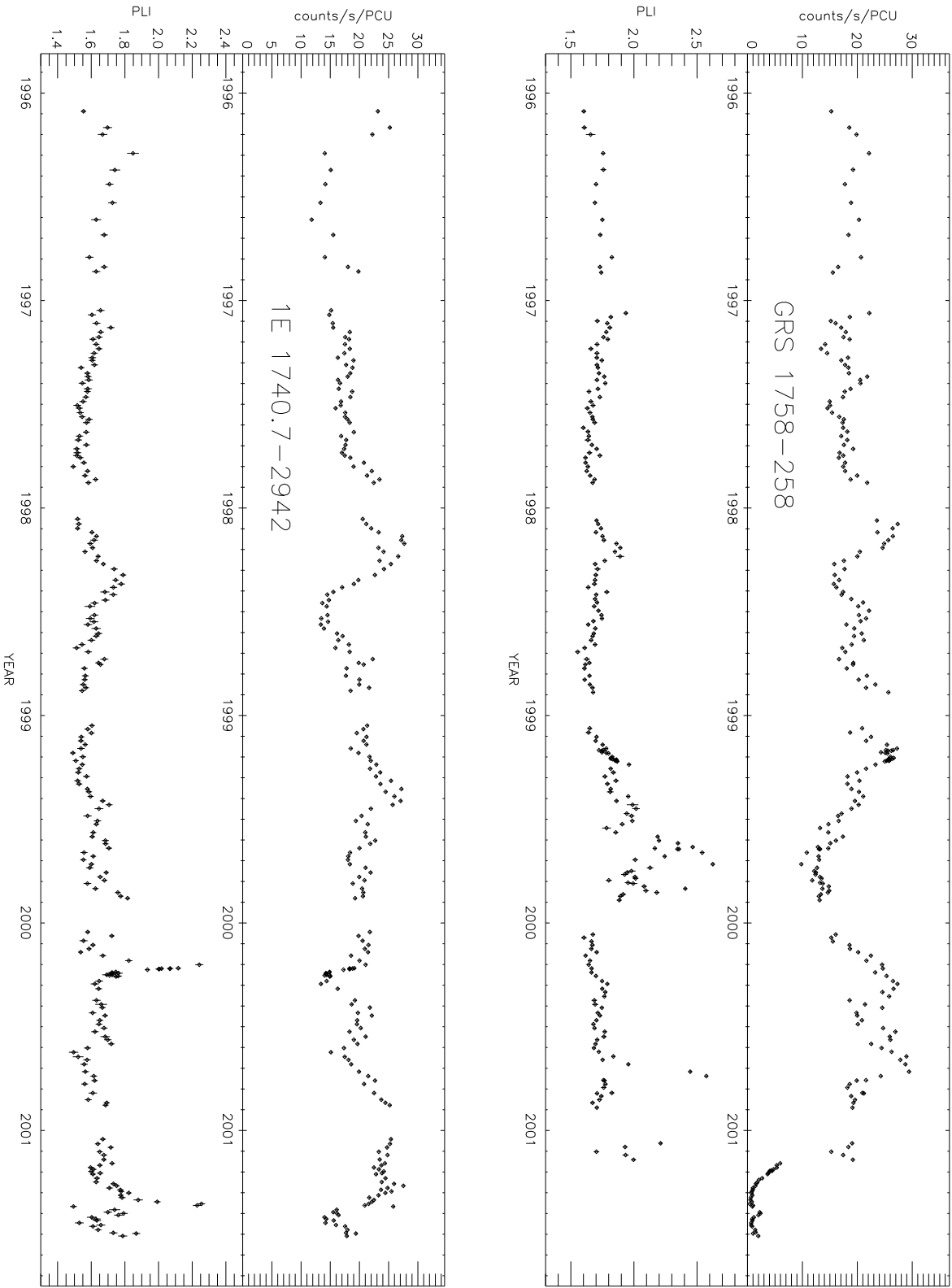}
\caption{Count rate and power law index from \grs\ and \onee\ as a function
of time from \it RXTE \rm monitoring data.}
\end{figure}

\cyg, \cygth, \onee, \grs, and \gx\ have been monitored by the \it
Rossi X-Ray Timing Explorer (RXTE)\rm\ since the start of science
operations in 1996 February.  \cyg, \cygth, and \gx\ are bright enough
to be monitored by the All-Sky Monitor (ASM) instrument, and we have
used quick-look results provided by the ASM team.  \onee\
and \grs, which are fainter and in a crowded part of the sky, have
been monitored by periodic visits with the highly-sensitive
Proportional Counter Array (PCA) instrument \citep{Ja96}.

The PCA monitoring observations, which last about 1500 s each,
occurred monthly during 1996, weekly until early this year, and are
now begin taken twice per week.  There are data gaps from November to
January every year due to a constraint against pointing the PCA near
the Sun.  In \citet[hereafter Paper I]{Ma99} and \citet{Sm97} we
discussed the details of our observing strategy for \onee\ and \grs,
including offset-pointing to avoid nearby sources and subtraction of
Galactic diffuse emission.  The principal changes since Paper I was
written are the new data acquired and a reanalysis of the whole data
set using updated PCA response and background models (the new faint
source model).

\begin{figure}
\plotone{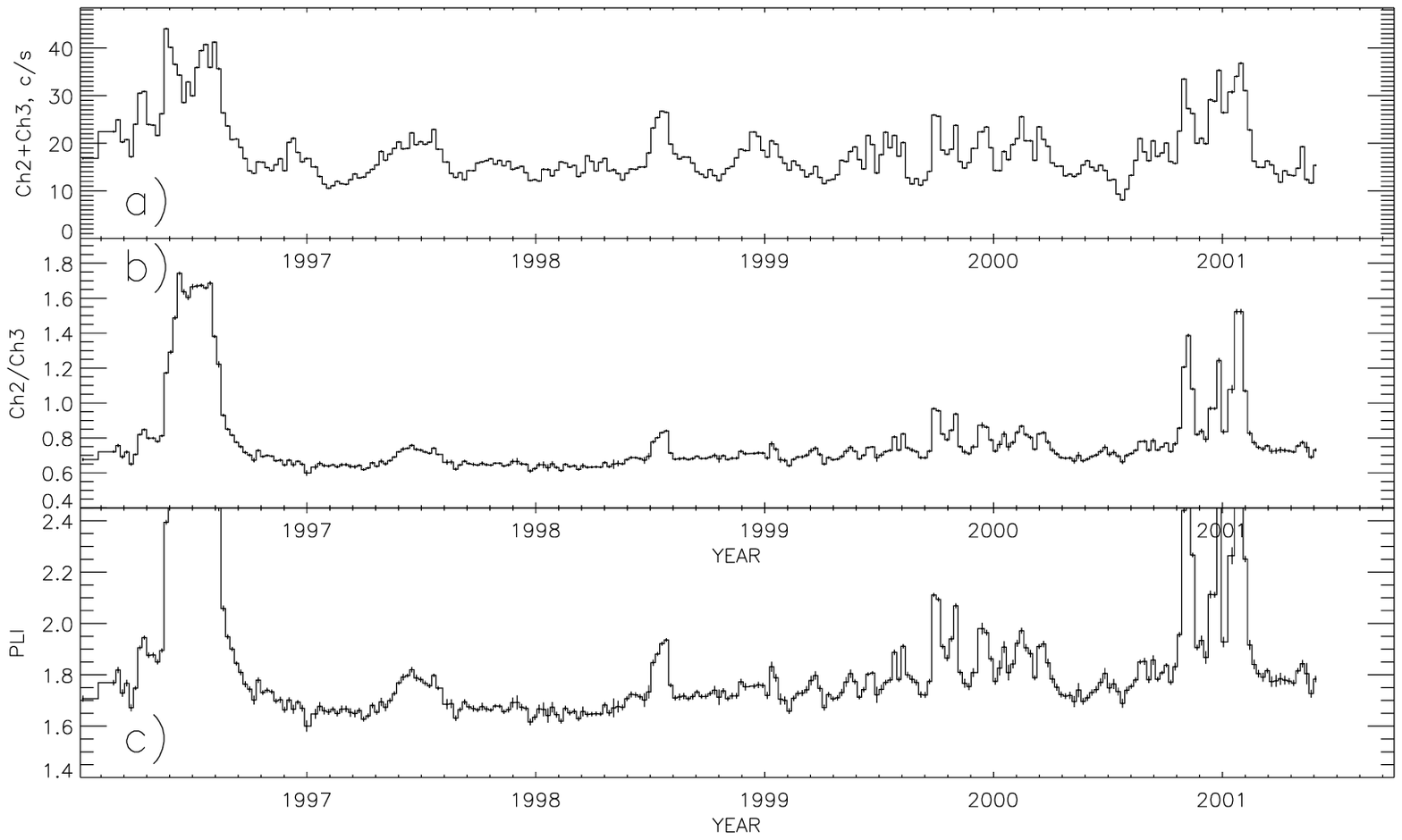}
\caption{Summed count rate in channels 2 and 3, ratio between
channels 2 and 3, and derived power law index in \cyg\ from
\it RXTE \rm ASM data.}
\end{figure}

\begin{figure}
\plotone{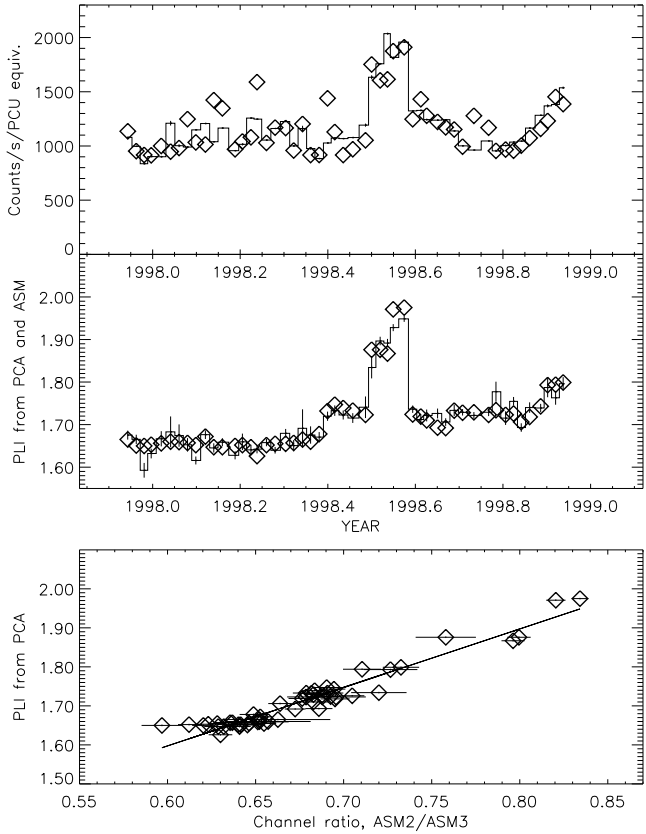}
\caption{ \it RXTE \rm ASM data (histograms) and PCA data (points) for \cyg\ during
1998.  Top: count rates.  The PCA data (diamonds) are in counts/second/PCU from 3-12 keV,
the ASM energy range.
The PCA error bars are much smaller than the plotting symbols. The ASM count rates (histogram)
have been normalized to
have the same average value as the PCA data.
Middle: Fitted PLI from the PCA data (diamonds) and equivalent index derived
from the ASM data by equation 1 (histogram).  The PCA error bars are comparable to the size of the
plotting symbol.
Bottom: The same data points as in the middle plot, but with the PCA PLI plotted against
the ASM count ratio, showing the linear fit (equation 1).}
\end{figure}

\begin{figure}
\plotone{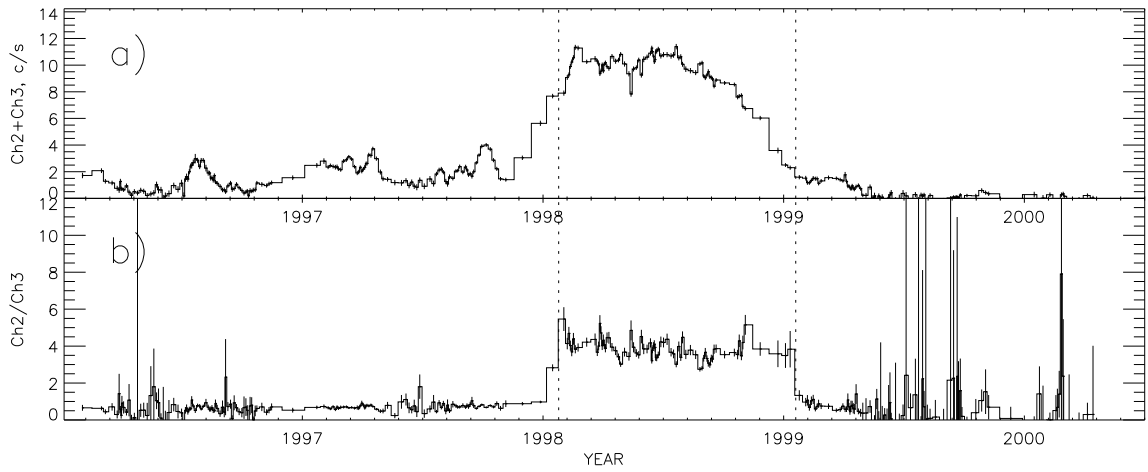}
\caption{Summed count rate in channels 2 and 3 and ratio between
channels 2 and 3 (spectral softness) in \gx\ from
\it RXTE \rm ASM data.  Dotted lines show where there is evidence
of a slight lag between flux and hardness variation at the
state transitions.}
\end{figure}

\begin{figure}
\plotone{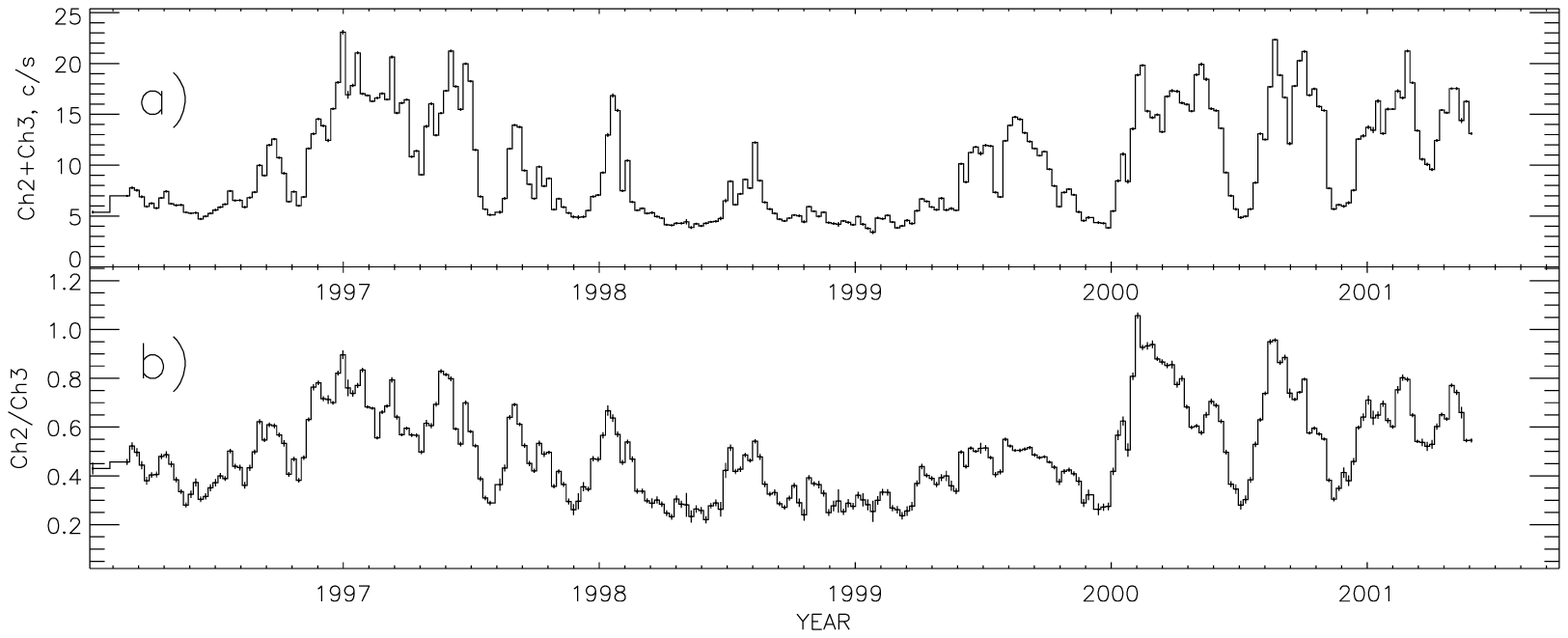}
\caption{Summed count rate in channels 2 and 3 and ratio between
channels 2 and 3 (spectral softness) in \cygth\ from
\it RXTE \rm ASM data.}
\end{figure}

Figure 1 shows the count rate and best-fit power-law index (PLI) in
photon space as a function of time for \onee\ and \grs.  The count
rate is in the range 2.5 - 25 keV, uncorrected for interstellar
absorption or instrument response.  The PLI was fitted to the data in
the same range.

Figure 2 shows ASM data for \cyg.  To restrict our attention to the
power-law component of the emission rather than the disk blackbody, we
sum only ASM channels 2 (3.0-5.0 keV) and 3 (5.0-12.1 keV).  To match
the \onee\ and \grs\ data as closely as possible, we plot weekly
accumulations of ASM data. Figure 2a shows the sum of the count rate
in these two channels as a measure of the source flux.  Figure
2b shows the ratio of these two channels.  

We used weekly PCA pointings to \cyg\ in 1998 to find the true PLI for
comparison with contemporaneous ASM data.  Figure 3 shows an expanded
view of the ASM data with the PCA data superimposed.  The count rates (Figure 3a)
are for ASM channels 2+3 and for the PCA data restricted to the same
range (3.0-12.1 keV).
The PCA data show more variation not because of
statistical errors, which are far smaller than the plotting symbols
used, but because they represent a single (3000 s) snapshot instead of
the average of a series of snapshots taken over a week, as is the
case with the ASM data.  This indicates a significant
power at frequencies between about 1 and 300 $\mu$Hz, an interesting
phenomenon beyond the scope of this paper.

Figure 3b compares the PCA PLI with the best-fit linear transform of
the ASM channel ratio using this data set.  This relation between the
PLI and the ratio, $R$, of ASM channels 2 and 3 (Figure 2b) is 
\begin{equation}
\rm{PLI} = 1.499\rm{R} + 0.698.  
\end{equation}
Figure 3c shows the PLI from the PCA data plotted against the ASM
channel ratio for this data set, and the fit that gives equation 1.
We do not expect that this relation applies
for the softest data of Figure 2, where the disk blackbody emission
could dominate channel 2, since this did not occur in 1998.  The
derived PLI, shown in Figure 2c, should therefore not be considered
accurate when it is softer than about 2.1.

Figure 4 shows the ASM count rate and channel ratio for \gx.  Since
this source is much fainter than \cyg, and sometimes drops to
undetectable levels, the data are poorer.  The ASM coverage of any
given point in the sky is irregular, so when the data are accumulated
on a weekly basis as in Figure 2, there are periods where the
statistics are so poor that a plot of the data is difficult to
interpret.  We therefore produced Figure 4 by accumulating sets of 50
consecutive ASM snapshots and averaging them together.  One clear
state transition can be see in 1998.  Unfortunately, the transitions
into and out of the soft state both occurred around January, when the
\it RXTE \rm pointing schedule, driven by a Sun-angle constraint,
results annually in very poor coverage of the Galactic Center region
by the ASM.

Finally, Figure 5 shows the ASM rate and ratio data for \cygth, with
one-week accumulations as in Figure 2.  The hard x-ray spectra of this
source are more complicated than the others, including variable
absorption and a very bright iron line, presumably due to the effects
of a very dense wind from its companion \citep[e.g.,][]{Na93}.  We have
therefore not attempted to convert the channel 2 to channel 3 ratio
into a power-law index, but we leave it as the measure of softness
for this source.

\subsection{Correlating Photon Flux and Spectral Index}

\begin{figure}
\plotone{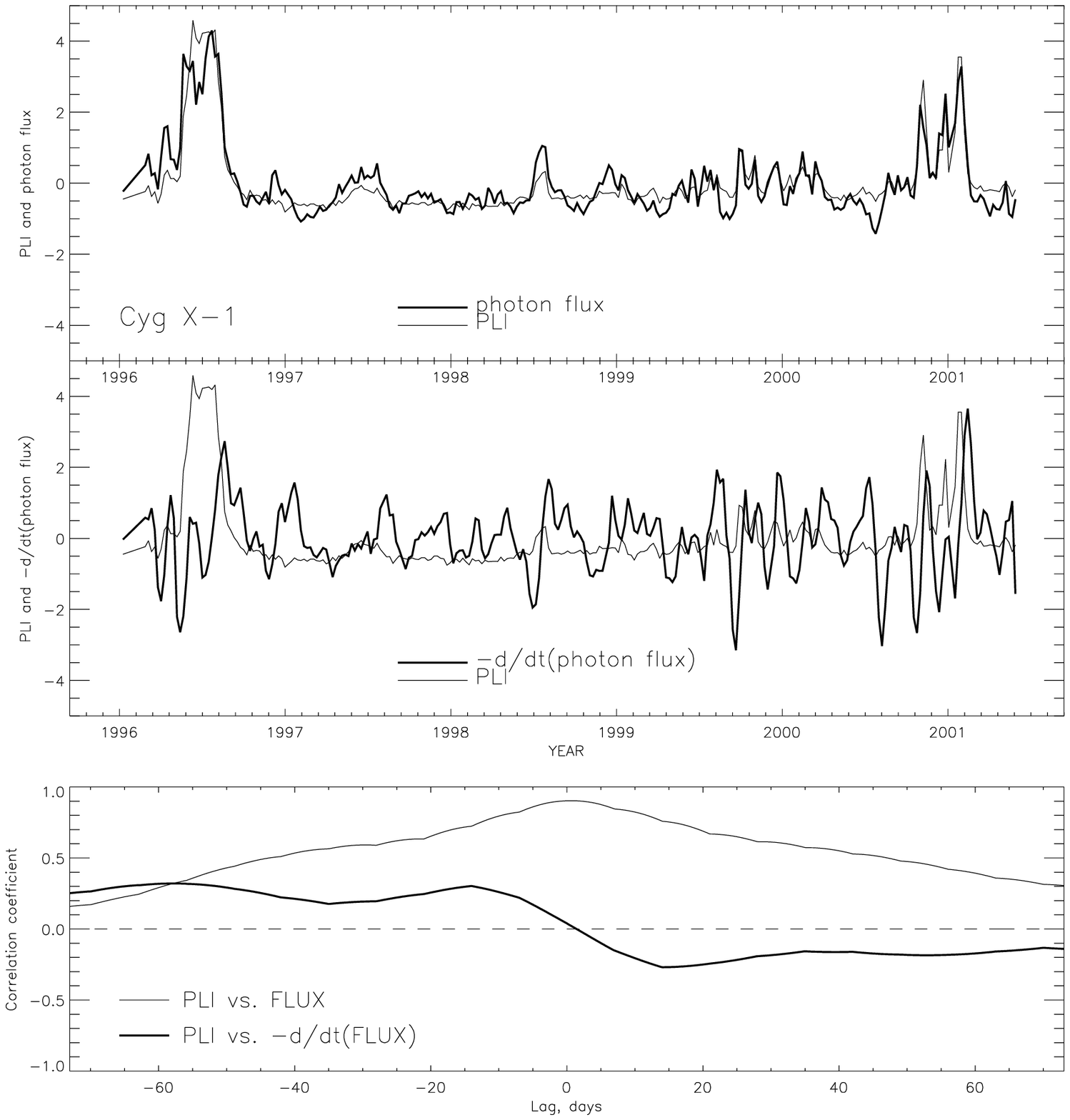}
\caption{Top panel: Scaled curves of PLI and photon flux for \cyg\ (see text).
Middle panel: Scaled curves of PLI and minus the time derivative of the
photon flux.  Bottom panel: correlation coefficient vs. lag for each of the
two sets of curves.}
\end{figure}

\begin{figure}
\plotone{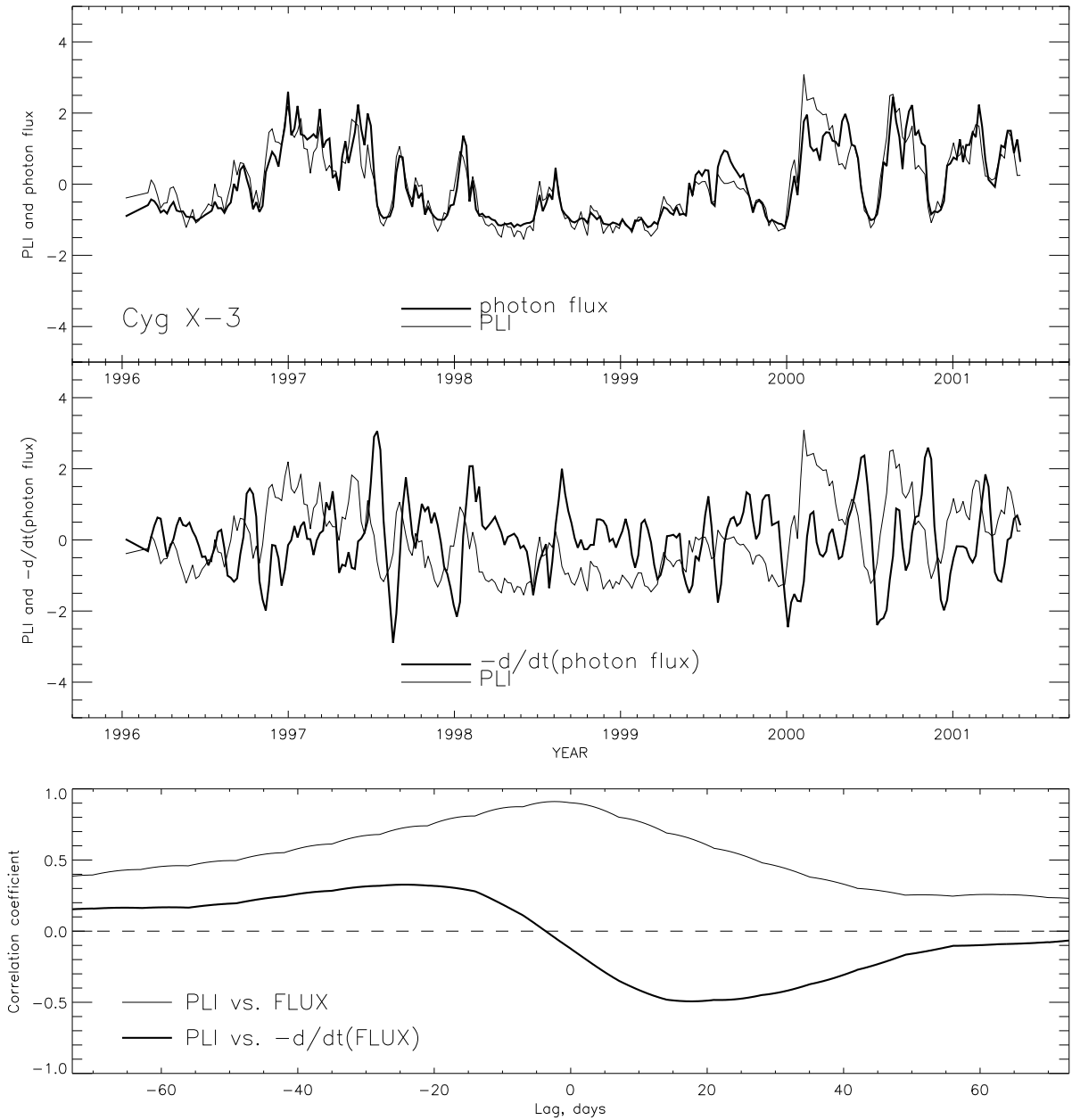}
\caption{Same as Figure 6 but for \cygth, with the count rate ratio of
ASM channel 2 to ASM channel 3 in place of the PLI.}
\end{figure}

\begin{figure}
\plotone{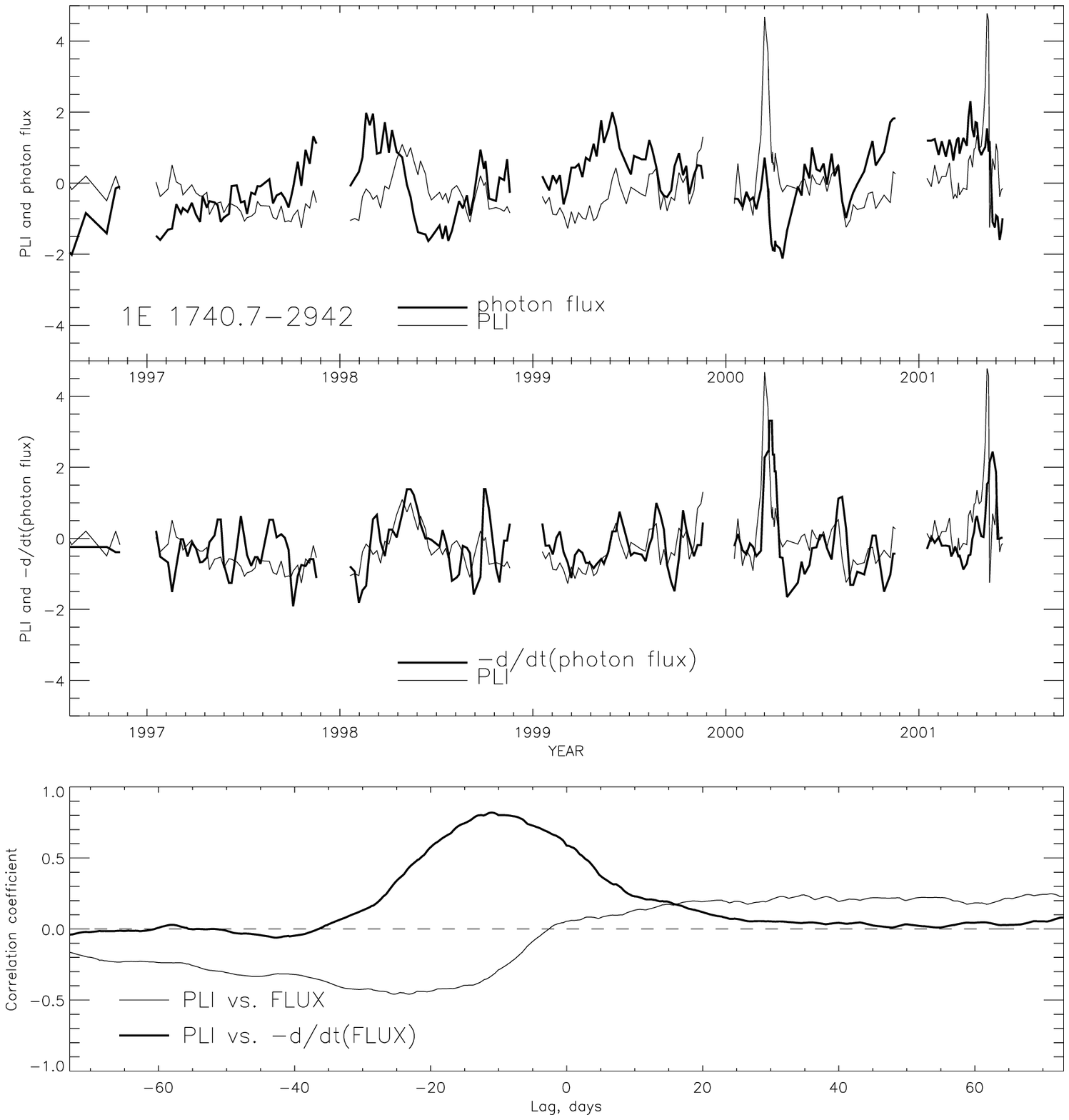}
\caption{Same as Figure 6 but for \onee.}
\end{figure}

\begin{figure}
\plotone{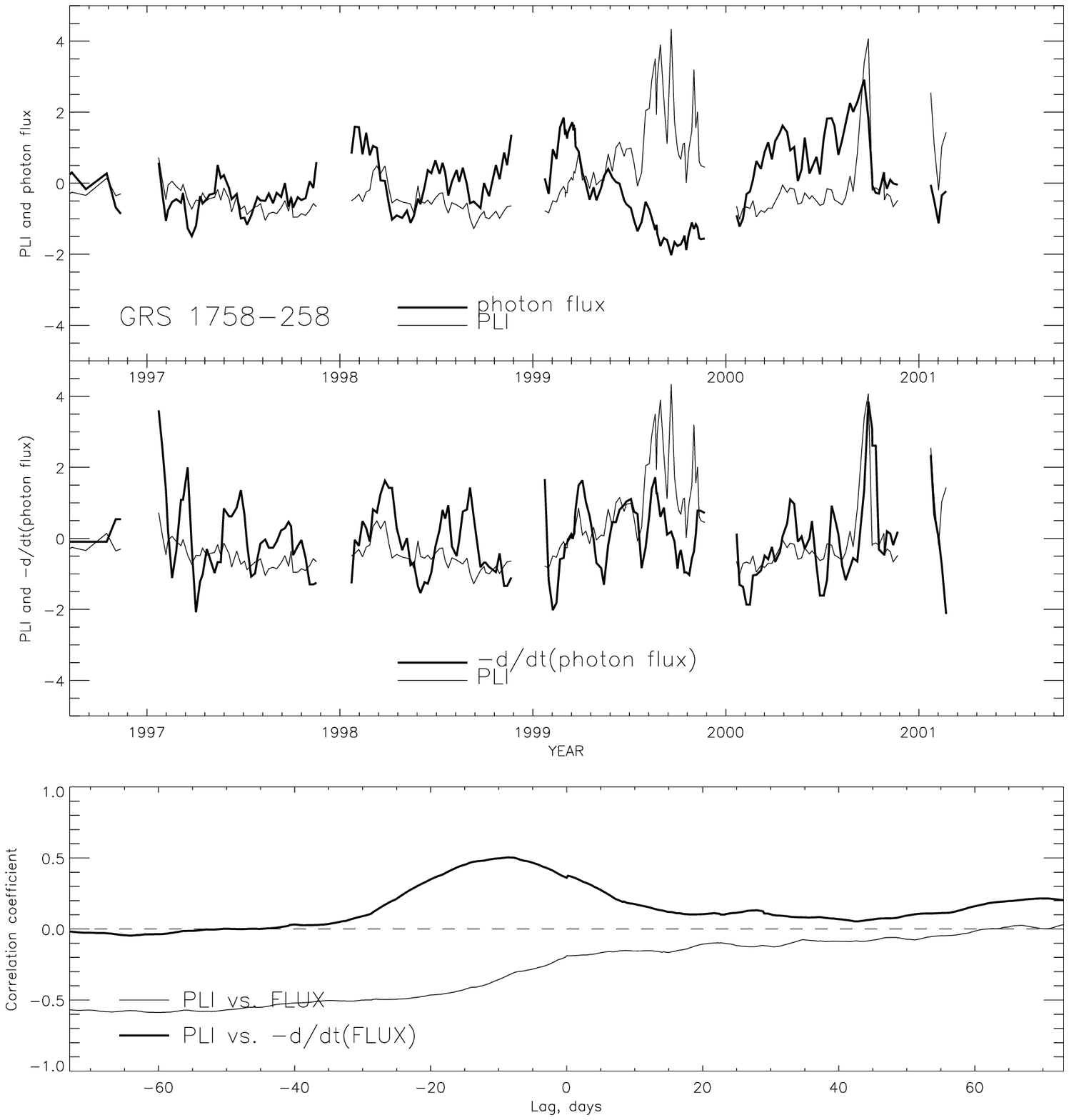}
\caption{Same as Figure 6 but for \grs.}
\end{figure}

There is a strong contrast in the behaviors shown in Figures 1 and 2.
In \cyg, the curves of brightening and softening appear almost
identical, while they have a very different time-dependence in \grs\
and \onee.  This phenomenon is the principal result of this paper.  In
order to make the relations between these quantities clearer in all
three sources, we have derived Figures 6-9 from the raw data in
Figures 1 and 2.  These figures compare the PLI with both the photon
flux and the time derivative of the photon flux for each source.  In
order to produce a set of standard comparisons, the count rates have
had the instrumental responses and interstellar absorption removed and
they are extrapolated to the range 2-100 keV using the fitted PLI.
This range was chosen to represent the total photon flux in the power
law part of the spectrum of each source.  Above 100 keV the photon
flux is small for both hard and soft state spectra, and below 100 keV
the error introduced in the hard state by ignoring the exponential
rolloff is also small.  The flux derivatives were calculated at a
given time by fitting a line to all the data within 15 days of that
point and using the slope.  For visual comparison, the PLI,
flux, and flux derivative curves are all scaled to have a mean of zero
and an rms of one.

\begin{deluxetable}{ccllllllll}
\tabletypesize{\scriptsize}
\tablecolumns{8} 
\tablewidth{0pc} 
\tablecaption{Cross-correlation peak values and peak lags, for pairs of
quantities $Q1$ and $Q2$}
\tablehead{ 
\colhead{$Q1$} & \colhead{$Q2$} & 
\multicolumn{2}{c}{\cyg } &
\multicolumn{2}{c}{\cygth } &
\multicolumn{2}{c}{\onee} &\multicolumn{2}{c}{\grs} \\
\cline{3-4} \cline{5-6} \cline{7-8} \cline{9-10}\\
\colhead{} & \colhead{} & \colhead{Value} & \colhead{Lag (dy)} &
\colhead{Value} & \colhead{Lag (dy)} &
\colhead{Value} & \colhead{Lag (dy)} &
\colhead{Value} & \colhead{Lag (dy)} }
\startdata
$\gamma$\tablenotemark{a} & $C$\tablenotemark{b}    &
             0.803$\pm$0.016 & -1.2$\pm$2.0     &                   
             0.906$\pm$0.016 & -2.9$\pm$3.3  &
                -0.426$\pm$0.026 & -25.45$\pm$0.15 &
                -0.596$\pm$0.020 & -60.2$\pm$6.6 \\
$\gamma$ & $F$\tablenotemark{c}  & 
             0.871$\pm$0.014 & 0.3$\pm$1.9  &   
             \nodata &            \nodata &
            -0.459$\pm$0.038  & -25.4$\pm$2.5 &
            -0.587$\pm$0.026  & -60.0$\pm$ 8.0 \\
$\gamma$ & $E$\tablenotemark{d}  &
             0.874$\pm$0.013 & 0.0$\pm$2.4 &                          
             \nodata &            \nodata &
               -0.632$\pm$0.037 & -11.3$\pm$2.1  &  
               -0.712$\pm$0.021 &-24$\pm$22 \\
$\gamma$ & $-dC/dt$ &  0.327$\pm$0.028&-13.4$\pm$1.2 &
                       -0.471$\pm$0.026&19.2$\pm$3.9  &
                      0.752$\pm$0.014 & -11.00$\pm$0.10 &       
                      0.460$\pm$0.012 &-7.85$\pm$ 0.60 \\
$\gamma$ & $-dF/dt$ &        0.363$\pm$0.028&-13.40$\pm$0.60 &
             \nodata &            \nodata &
                    0.820$\pm$0.043 & -11.05$\pm$0.20      &
                    0.504$\pm$0.039 &-8.50$\pm$ 0.65 \\
$\gamma$ & $-dE/dt$ &  0.215$\pm$0.043&-13.9$\pm$4.1   & 
             \nodata &            \nodata &
                    0.591$\pm$0.085  & 10.8$\pm$3.6      &
                    0.444$\pm$0.061 &11.55$\pm$0.60 \\
\tableline
$\tau$\tablenotemark{e} & $C$ &    -0.838$\pm$0.021&0.3$\pm$1.7 &
             \nodata &            \nodata &
                      0.426$\pm$0.031 & -25.55$\pm$0.15 &
                      0.616$\pm$0.022&-60.0$\pm$1.7 \\
$\tau$ & $F$ &  -0.812$\pm$0.027&0.5$\pm$2.6   &
             \nodata &            \nodata &
                0.453$\pm$0.045 & -25.6$\pm$2.1 &
                 0.647$\pm$0.018 & -58.6$\pm$ 5.4 \\
$\tau$ & $E$ & -0.933$\pm$0.014 &0.0$\pm$2.1 &
             \nodata &            \nodata &
                0.632$\pm$0.037 & -11.3$\pm$8.7 &
                0.734$\pm$0.022 & -24$\pm$21 \\
$\tau$ & $dC/dt$ & 0.360$\pm$0.033&-13.4$\pm$3.5 &
             \nodata &            \nodata &
              0.748$\pm$0.017 & -11.00$\pm$0.10      &
              0.457$\pm$0.014 & -8.45$\pm$0.60\\
$\tau$ & $dF/dt$ & 0.387$\pm$0.033&-13.4$\pm$2.6 &
             \nodata &            \nodata &
                          0.777$\pm$0.044 & -11.2$\pm$2.3 &
                          0.525$\pm$0.044 & -10.8$\pm$1.0 \\
$\tau$ & $dE/dt$ &  0.166$\pm$0.035&9$\pm$19 &
             \nodata &            \nodata &
                            0.572$\pm$0.089 & 10.8$\pm$6.8 &
                            0.424$\pm$0.060 & 11.4$\pm$2.8 \\
\enddata 
\tablecomments{For \cyg, $\gamma$, $\tau$, $E$, and $F$ are derived based
on the relation between $\gamma$ and the ASM channel ratio
(see Equation 1). ``$\gamma$'' for \cygth\ is replaced by the
raw ASM channel ratio, and $\tau$, $E$, and $F$ cannot be derived,
since the spectral shape is complex.}
\tablenotetext{a}{Power-law index.}
\tablenotetext{b}{Raw count rate (PCA 2.5-25 keV or ASM channels 2+3).}
\tablenotetext{c}{Flux, photons cm$^{-2}$s$^{-1}$, 2-100 keV.}
\tablenotetext{d}{Energy flux, ergs cm$^{-2}$s$^{-1}$, 2-100 keV.}
\tablenotetext{e}{Optical depth derived from $\gamma$ (see $\S$3.3).}

\end{deluxetable}

The last panel of each figure shows the cross-correlation between the
curves in each of the two panels above it.  The peak correlations
(positive or negative) and corresponding lags are shown in Table 1.  A
positive lag is defined as the PLI lagging behind the other quantity.
To cross-correlate the unevenly sampled data, we interpolated between
points on the PLI curve in order to get values at times corresponding
to the shifted flux and flux derivative data.  

The errors in the values of the correlation coefficient and the
best-fit lag were estimated empirically.  All points in the flux and
PLI curves were randomly perturbed according to a Gaussian
distribution with a width determined by the errors of the
measurements.  The analysis process, including calculation of the flux
derivative and cross-correlation, was then repeated for the perturbed
data set.  This was done 1000 times with different random
perturbations for each cross-correlation.  The errors quoted on the
correlation coefficients and the best-fit lags are the intervals that
enclosed 95\% of the results from the perturbed cases.

To make the results in Table 1 as mutually comparable as possible, we
averaged the ASM data for \cyg\ and \cygth\ into time intervals
centered on the times of each of the PCA pointings to \onee, so that
the number of data points and irregularity of sampling is similar for
all four sources.  Figures 6, 7 and later figures for \cyg\ and
\cygth\ use regular weekly binning of the data for display (as do
Figures 2 and 5).  Using these weekly bins or using daily bins, the
cross-correlation results are still similar to Table 1.  For \cygth\
the ratio between ASM channels 2 and 3 is used as a surrogate for the
PLI.  For \cyg\ the PLI and ASM channel ratio were shown to be
linearly correlated (Figure 3), and therefore they yield the same
curve when scaled to zero mean and unit rms.

\cyg\ and \cygth\ (Figures 6 and 7) show an extremely good correlation
of the PLI with the photon flux or count rate, but no clear
correlation with its derivative.  \onee\ (Figure 8, Table 1) shows the
opposite behavior: a very good correlation of the PLI with the
derivative of the photon flux, and a poor correlation with the flux
itself.  The PLI leads the flux derivative significantly, i.e. the
beginning of an episode of softening is a predictor of a drop in the
photon flux. \grs\ (Figure 9) shows behavior more like that of \onee\
than that of \cyg, although the correlation with the flux derivative
is not as good.

\begin{figure}
\plotone{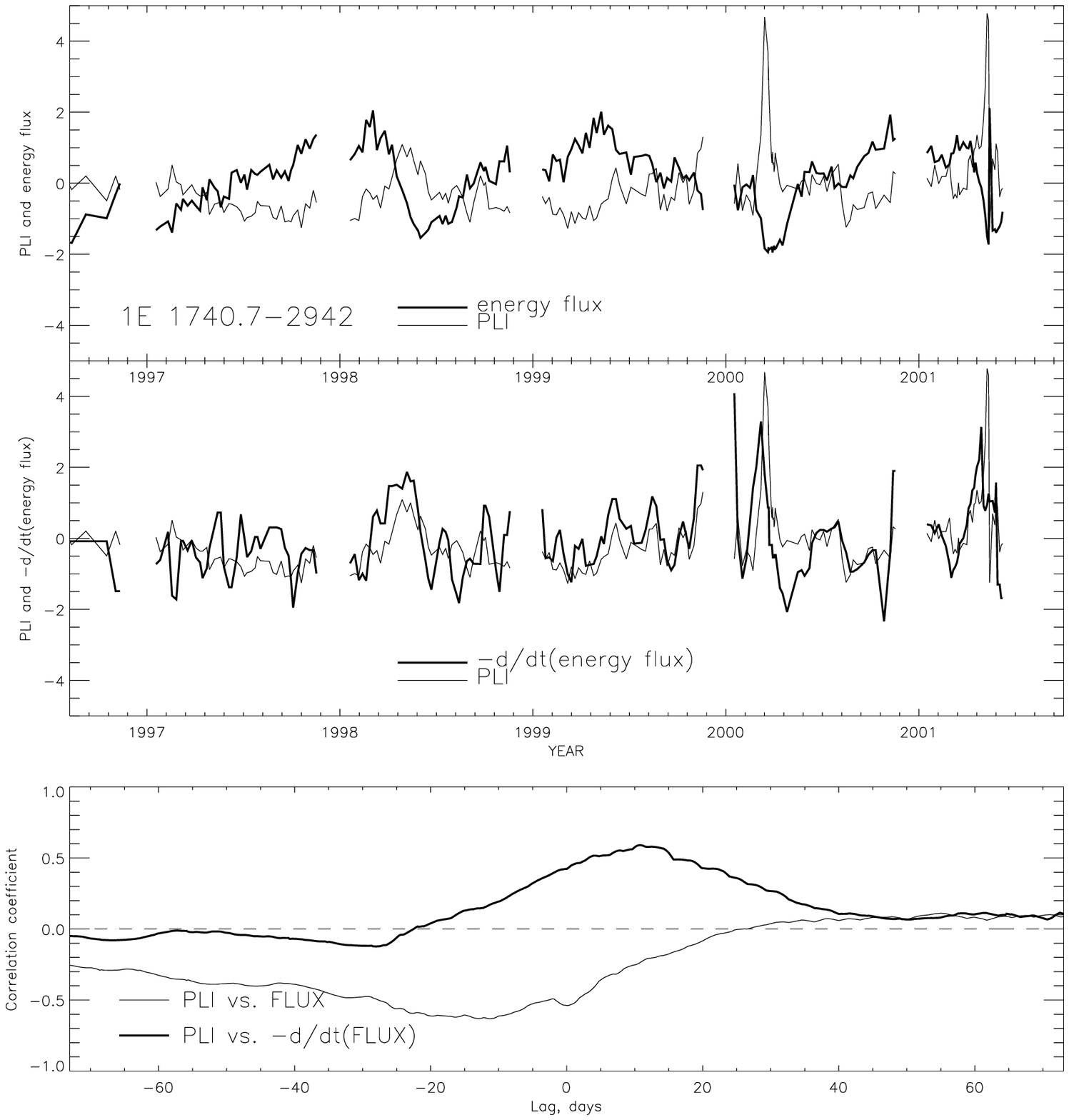}
\caption{Same as Figure 8 but with energy flux (2-100 keV) in place
of photon flux.}
\end{figure}

\begin{figure}
\plotone{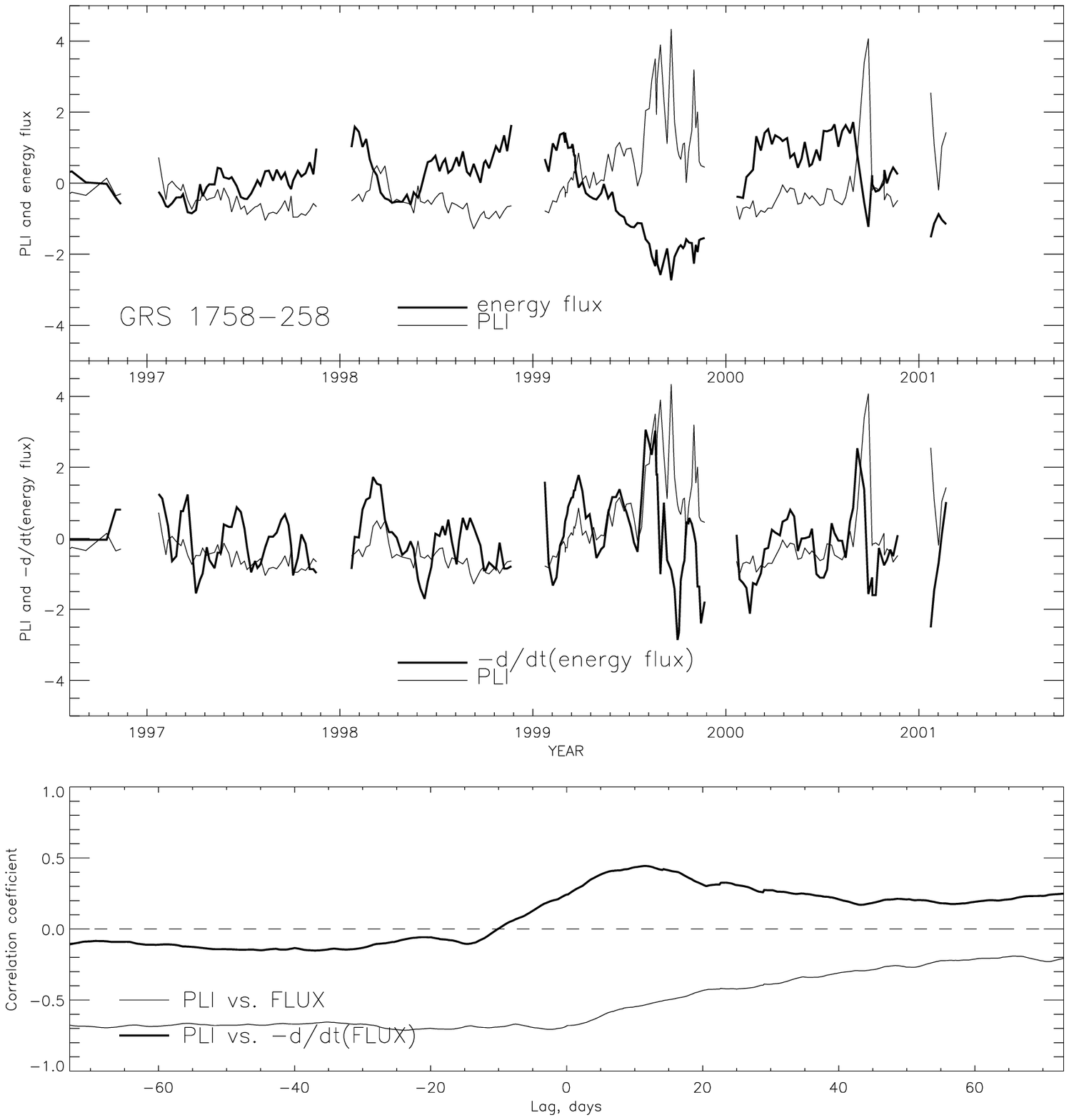}
\caption{Same as Figure 9 but with energy flux (2-100 keV) in place
of photon flux.}
\end{figure}

Qualitatively, the results of Figures 6-9 are the same whether the
quantity compared to the PLI is the 2-100 keV photon flux as shown or
the raw count rate in the instrument (PCA or ASM; see Table 1).  When
the energy flux from 2-100 keV is used in place of the photon flux,
\onee\ and \grs\ still show a better correlation with the flux
derivative than with the flux, but the correlations are weaker and the
sign of the lag is reversed: the energy flux derivative leads the PLI
(Figures 10 and 11).  The corresponding plot for \cyg\ using the
energy flux looks similar to the photon flux plot (Figure 6).  Because
we don't have a formula like equation 1 to convert the \cygth\ ASM
channel ratio to a PLI, we don't calculate the energy flux for \cygth.

We don't calculate a correlation for \gx\ because there is only a
single major episode in the plot shown in Figure 4: the rise to the
soft state around January 1998 and the fall back to the hard state
around December 1998.  The first dotted line in Figure 4 shows the
time at which the power law becomes soft.  The second dotted line
shows the time at which it hardens again.  Note that at this time the
flux has already dropped to a level considerably lower than that at
which the power law first softened.  Although this is a very small
amount of data, qualitatively we might say that \gx\ resembles \cyg\
during the rise (softening correlated with flux) and \onee\ and \grs\
during the decline (a soft spectrum while the flux decreases).  It has
been noted that some black-hole x-ray nova outbursts have a
hard-to-soft transition at high luminosity followed by a soft-to-hard
transition at much lower luminosity \citep{Mi95}.  In
addition to citing the well-studied x-ray nova GS~1124-683 (Nova
Muscae 1991), Miyamoto et al. found evidence for a non-simultaneous change of
flux and PLI in \gx\ using data taken by several instruments in
1988 and 1991.  The total luminosity in the power law spectral
component peaked as the PLI was making a transition from -2 to -3.

\subsection{The most sudden changes}

Although most of the large variations in count rate and PLI in \onee\
and \grs\ have been well sampled by our observing program (see Figure
1), each source has on at least one occasion changed too rapidly for us to
follow the change in detail.  Each of these events is worth special
consideration, since the strongest tests of theoretical models for
these sources may come from the most extreme variations.

Near the end of 2001 February, the power law flux from \grs\ dropped
by an order of magnitude between one pointing and the next, leaving
behind a spectrum dominated by a weak blackbody component.  Based on a
model of two accretion flows (a sub-Keplerian halo and a thin disk),
we predicted \citep{Sm01c} that mass input from the companion had
ceased, and that the blackbody should decay smoothly over the
following weeks.  It did so, with a time constant of roughly 28 dy
\citep{Sm01a}.  Both the sharp drop (due to the abrupt loss of the
power-law component) and the subsequent decay of the soft flux are
visible in the count rate plot of Figure 1.  This decaying state was
similar to the usual black hole ``soft'' or ``high'' state, both
because it was dominated by thermal emission and because there was no
detectable fast x-ray variability.  A low-luminosity soft state for
this source was previously inferred by \citet{Gr97} from data taken
between 1991 Fall and 1992 Spring: during this period, \it GRANAT \rm
found only upper limits above 3 keV, while \it ROSAT \rm made a good
spectral measurement from 1-2.4 keV.  The combined data sets imply
that the spectrum had to be both faint and soft.  These results differ
from the canonical ``high/soft'' state in that the soft spectrum
appears at much lower luminosity than the hard state.  The hard to
soft transition is usually said to result from an increase in
accretion rate and usually involves an increase in luminosity.

\begin{figure}
\plotone{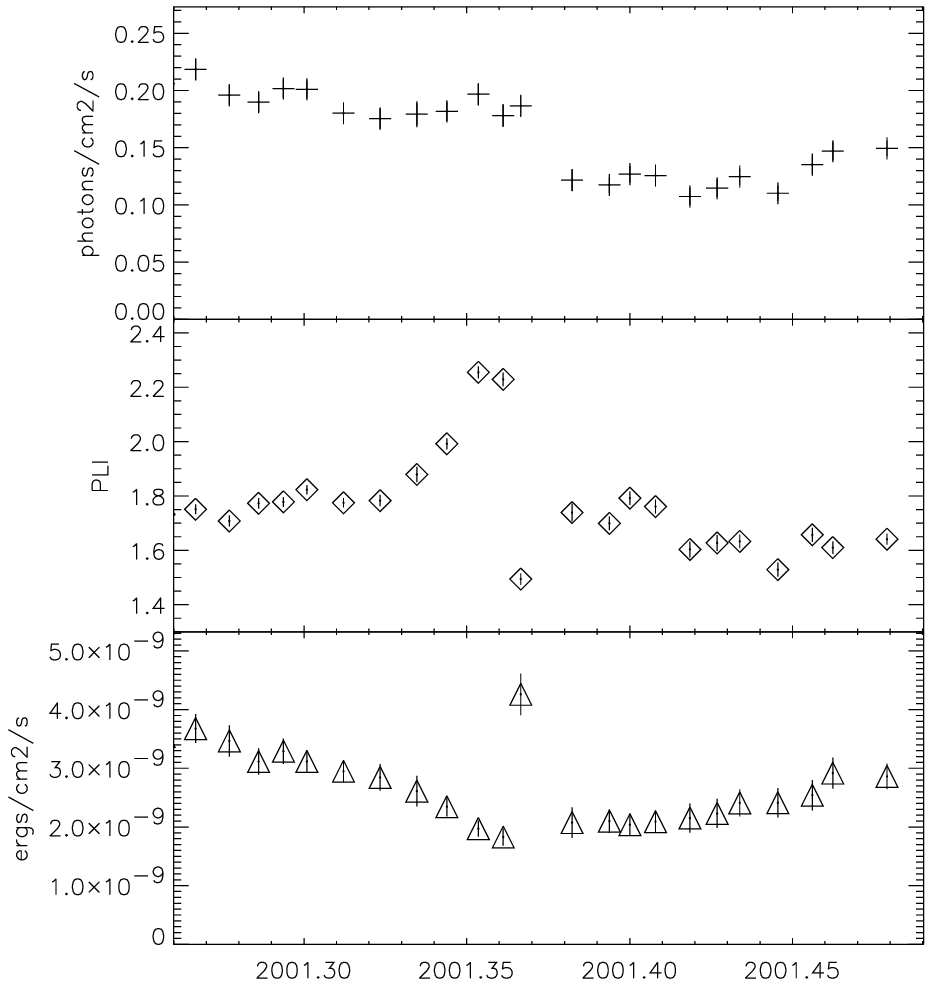}
\caption{Evolution of photon flux (2-100 keV), PLI, and energy flux
(2-100 keV) as a function of time around the episode of sudden hardening
in \onee\ on 2001 May 14.}
\end{figure}

A single pointing on 2001 May 14 found that \onee\ had suddenly jumped
from one of softest PLIs ever seen from this source (2.23 $\pm$ 0.02) to
one of the hardest (1.49 $\pm$ 0.02) \citep{Sm01b}.  As can be seen in
Figure 12, the 2-100 keV (extrapolated) photon flux remained virtually
unchanged while the spectrum hardened, so that the 2-100 keV energy
flux briefly doubled, returning at the next pointing to its
previous value.  The raw PCA count rate (2.5-25 keV) increased
slightly (this can be seen in Figure 1). If this one point was taken
out of the data, the behavior in this period would be very
similar to what we observe at other times: a softening preceeding a
drop in the photon flux by about 10 dy (see above).  We can be fairly
certain that the May 14 result is not due to a second,
very hard source turning on in the PCA field of view, because the
hardening involved not just an increase in high-energy counts, but
also a decrease in low-energy counts.  Although this may be
a rare sort of event, it might also be common but short-lived,
and detected only once for that reason.

\section{Discussion}

\subsection{Basic conclusions}

\cyg\ has long been the prototype which theories of black-hole
accretion have tried to model.  The results above make it clear that
there can be a richer variety of behavior even in other persistent
black-hole binaries than is observed in \cyg.  Specifically, any model
in which a single variable (e.g. a single accretion rate) controls
both the luminosity and the spectrum is clearly inadequate to explain
the behavior of \onee\ and \grs, even if it is extremely successful with
\cyg.

The clearest contrast in behavior is between \cyg\ and \onee:  the
PLI correlates with the photon flux in the former and its derivative
in the latter.  \cygth\ is very much like \cyg\ in this regard,
and \grs\ is more like \onee.  More data are clearly needed to
place \gx\ firmly in either category.  In $\S$1, we
reviewed the status of knowledge about the companion of each
source.  For \grs, \cyg, and \cygth, we can say that the sources
with probable high-mass companions (wind accretors) have the
PLI tracking with the flux, and the source with a probable low-mass
companion has the PLI tracking the flux derivative.  The situation
is less clear for \gx\ and \onee, in the former case because there
are not enough data on the x-ray behavior, and in the latter case because
there is neither a convincing measurement or upper limit for the
companion.  

If the correlation between companion type and x-ray behavior holds,
the size of the accretion disk may be what distinguishes the two sorts
of x-ray behavior.  Wind accretors are expected to have a very small
specific angular momentum in the accreting material, and therefore
should form a very small disk.  The complicated behavior in \onee\ and
\grs\ might then be related to viscous propagation delays that are
much smaller in \cyg.  For a constant coefficient of kinematic
viscosity, $\nu$, the viscous timescale in a standard thin disk
$t_{\rm{visc}} \sim R^2/\nu$ \citep{Fr92}, so the accretion disk of
\cyg\ would only have to be an order of magnitude smaller than that of
the LMXRBs to have a timescale $<$ 1 dy; i.e. it could still be over a
thousand gravitational radii ($R_g$) in size.  One recent model of
wind-fed accretion \citep{Be01} produces a standard hard-state
spectrum with a very quickly accreting, ``inviscid'' disk that is
extremely small: only 14$R_g$.  Compact objects which accrete via
Roche-lobe overflow, however, have disks of a size on the order of the
binary separation.  These large disks have a viscous timescale
(spiraling in of matter) which can be on the order of days to months
\citep{Fr92}, and could therefore be related to the time lags between
correlated quantities in Figures 8 and 9.  The $\sim$ 28-dy decay of
\grs\ in the soft state \citep{Sm01a} may be a direct measurement of
the viscous timescale in that source.

Another significant difference between \cyg\ on the one hand and
\onee\ and \grs\ on the other is that the latter sources have bright
radio lobes \citep{Mi92,Ro92}.  This could be a consequence of disk
size being related to jet collimation, of disruption of incipient
jets by the companion wind in \cyg, or of some other phenomenon.  If a
substantial fraction of the hard x-ray emission in \onee\ and \grs\
occurred in the jets instead of near the disk, one might suggest that
the time delays are due to the time required for information about
changes in the accretion rate to propagate along the jet to the
emitting region (e.g. a shock).  The spectrum observed in \grs\ as its
emission shut off early in 2001 \citep{Sm01a} eliminates this
possibility, however.  That spectrum was dominated by a soft thermal
component, brighter at first than any that was ever seen in the hard
state.  Since disk and jet spectral components would simply add,
the thermal flux cannot be the jet emission since it is not usually seen
in the hard state.

The hard-to-soft transition in \cyg\ has sometimes been explained in
the context of a central hot, advection-dominated flow with an outer
thin disk, the transition radius determined by the accretion rate
\citep[e.g.,][]{Es98,Ja00}.  As the accretion rate increases, the central hot
flow becomes unstable to collapse because it can cool itself more
efficiently by bremsstrahlung.  Here we will refer to this as a
``static'' soft state, since it can be reached by a gradual,
quasi-static change in the accretion rate.  In \onee\ and \grs, the
softest spectra observed are related to dropping photon fluxes (see
Figures 8 \& 9).  Because the spectral state is related to the rate of
change of luminosity, we refer to this as a ``dynamical'' soft state.
Clearly it cannot be related to high accretion rates;
\grs\ showed a soft spectrum down to a few
percent of its usual hard luminosity when it shut off \citep{Sm01a}.

\subsection{Relation to previous results}

In Paper I, we had only the episodes of brightening and softening
occurring in \onee\ and \grs\ in 1998 to look at.  From these data
alone, there was no way to tell whether the PLI was following the same
profile as the photon flux but with a delay, or whether it instead
followed the flux's derivative (these options would of course be
completely indistinguishable for sinusoidal variations).  It is the
newer data, including sharp drops in the count rate resulting in more
significant softenings, which have decided the question.

We suggested two qualitative models in the Paper I for
explaining the lack of simultaneity in the variations in 
flux and PLI.  Both pictures involved the simultaneous presence of
two accretion flows: a thin Keplerian disk and a hot, sub-Keplerian
halo (e.g. an advection-dominated flow \citep{Es98} or a shocked flow
\citep{Ch95}).

Our first picture followed a prediction of \citet{Ch95}:
the sub-Keplerian flow might exist at all radii, not just
within a transition radius, and if the mass accretion rate was boosted
to both components at once, the halo would brighten at all radii
almost instantaneously (near the free-fall timescale), while the inner
regions of the thin disk would only brighten after the extra mass had
wound its way in on the viscous timescale.  We suggested that when
this finally happened, the extra soft photons would cool the halo and
the spectrum would soften.

Our second picture followed a suggestion by \citet{Min96} related
to the soft-to-hard state transition in the decaying phase of soft
x-ray transients.  He suggested that, for accretion rates where both
thin-disk and advection-dominated flows are stable, the thin disk may
tend to persist once it is established, i.e. the evaporation back to
the advection-dominated flow is slow and inefficient.  Thus, even if
the brightening is quasi-static (i.e. the accretion rate changes
slowly compared to the viscous timescale), there can still be a soft
stage at the end of the outburst when the accretion rate is low again
but the thin disk persists. Most disk analyses concentrate on finding
stable solutions; the issue of transformations between stable states
is difficult and much less well-explored.

\cyg\ provides a perfect test case to distinguish between these two
pictures.  Since it is expected to have a very short viscous timescale
(see above), under the first picture (viscous delay), we would expect
a much shorter delay.  The second picture (the persistent disk)
depends only on a hypothetical property of the inner disk, which
should be roughly independent of the outer parts of the accretion
flow.  Thus, the simultaneity of brightening and softening in Figure 2
causes us to prefer the viscous delay picture \citep{Ch95} over the
slow-evaporation picture for the delay in \grs\ and \onee.

Since we now have considerably more data, showing greater excursions
in PLI, a deeper examination of the two-flow picture \citep{Ch95}
is warranted.

\subsection{Correlating photon flux and derived halo depth}

The hard component in black-hole-candidate spectra has long been
attributed to inverse Compton scattering of soft disk photons by
energetic electrons, usually but not always thermal.  These hot
electrons have been hypothesized to reside in several places, often a
hot, central region of the disk or a halo or corona above the disk.
The PLI of the Comptonized component is a function of the temperature
and optical depth of the Comptonizing plasma.  \citet{Pi95} have
modeled coronae above the disk and found that, for accretion rates
much less than Eddington, the halo temperature is nearly constant at a
value of about kT~=~100~keV even as the accretion rate varies by up to
four orders of magnitude.  A similar result was found by \citet{Ch95},
but the hot region was a shocked plasma, originating from a
sub-Keplerian flow, and occupying the radially innermost part of the
accretion flow.  The constant temperature holds as long as the
dominant cooling mechanism for the halo is the Comptonization of soft
photons and the accretion is significantly sub-Eddington.  The highest
luminosities derived from the \onee\ and \grs\ data of Figure 1 are
9.8 and 5.4~$\times 10^{-9}$ erg cm$^{-2}$ s$^{-1}$ (extrapolated to
2-100~keV and with interstellar absorption removed).  For a Galactic
Center distance of 8.5 kpc, this gives 8.5 and 4.7~$\times 10^{37}$
erg s$^{-1}$, or about 7\% and 4\% of the Eddington luminosity for a
10~M$_{\sun}$ black hole, so these objects could plausibly be in the
constant-temperature regime, in which the PLI depends only on the
optical depth $\tau$ of the scattering region.

\begin{figure}
\plotone{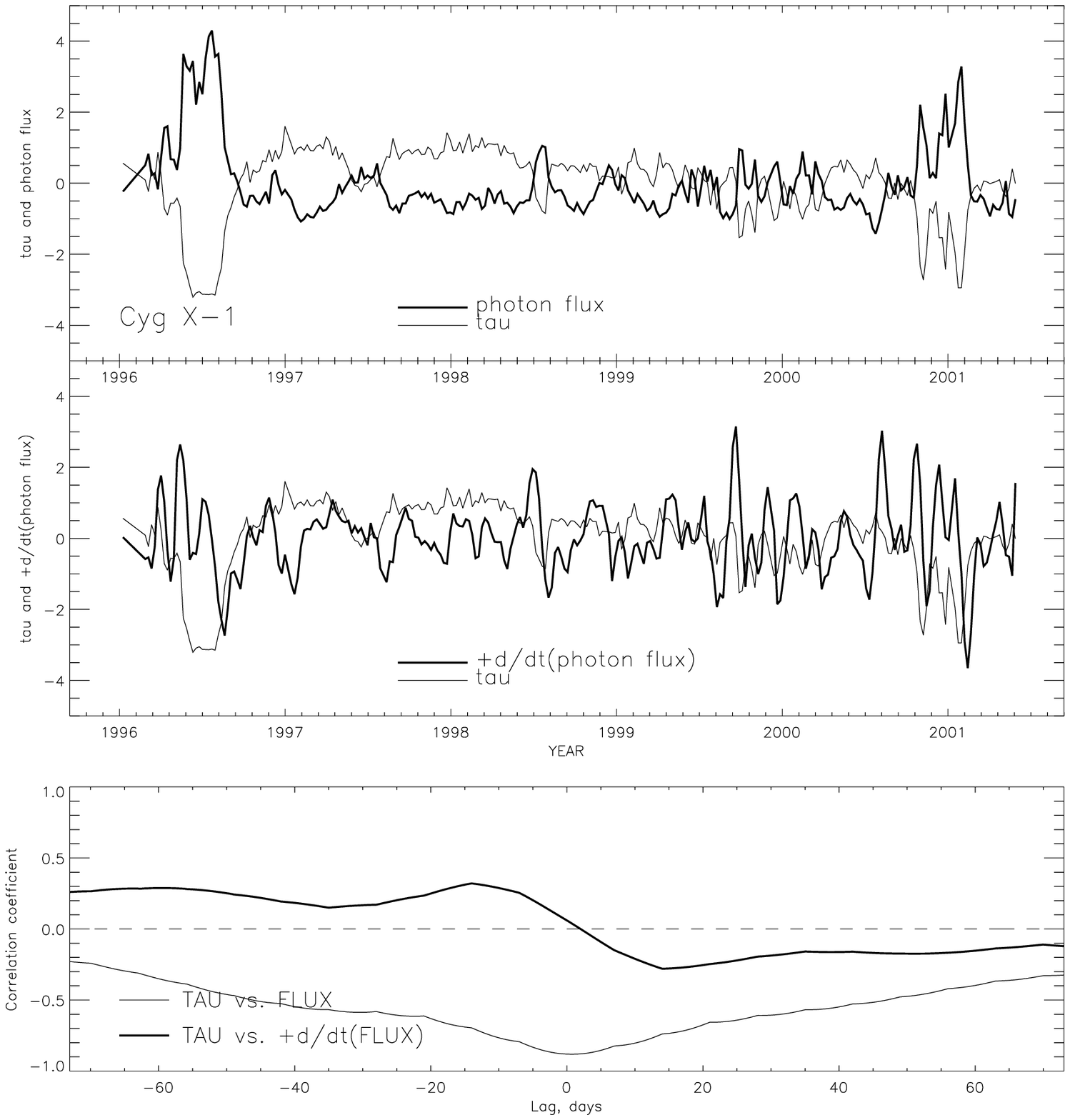}
\caption{Same as Figure 6 but with the PLI replaced by the derived
optical depth $\tau$ and the flux derivative shown with the
opposite sign.}
\end{figure}

\begin{figure}
\plotone{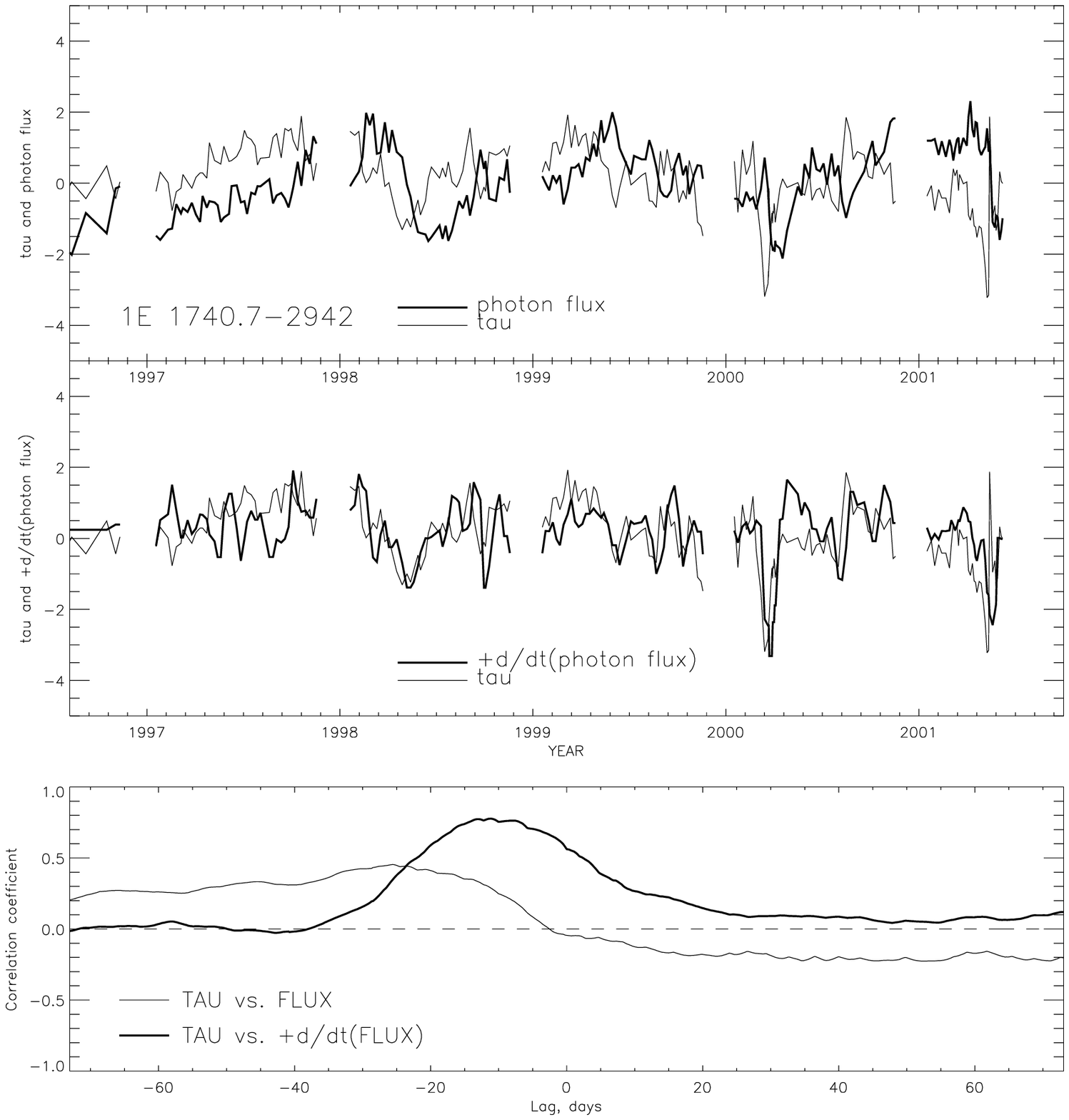}
\caption{Same as Figure 8 but with the PLI replaced by the derived
optical depth $\tau$ and the flux derivative shown with the
opposite sign.}
\end{figure}

\begin{figure}
\plotone{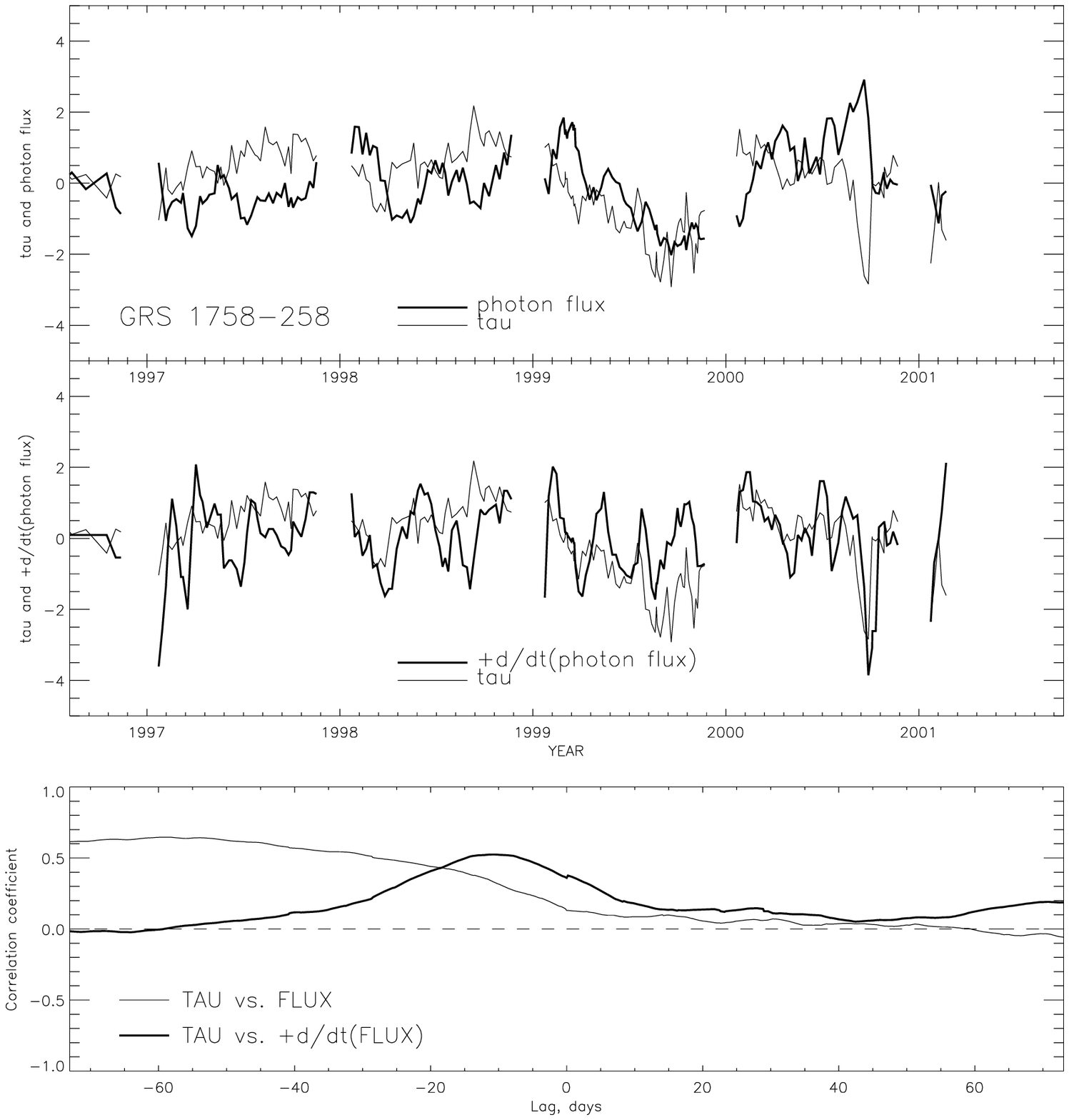}
\caption{Same as Figure 9 but with the PLI replaced by the derived
optical depth $\tau$ and the flux derivative shown with the
opposite sign.}
\end{figure}

\begin{figure}
\plotone{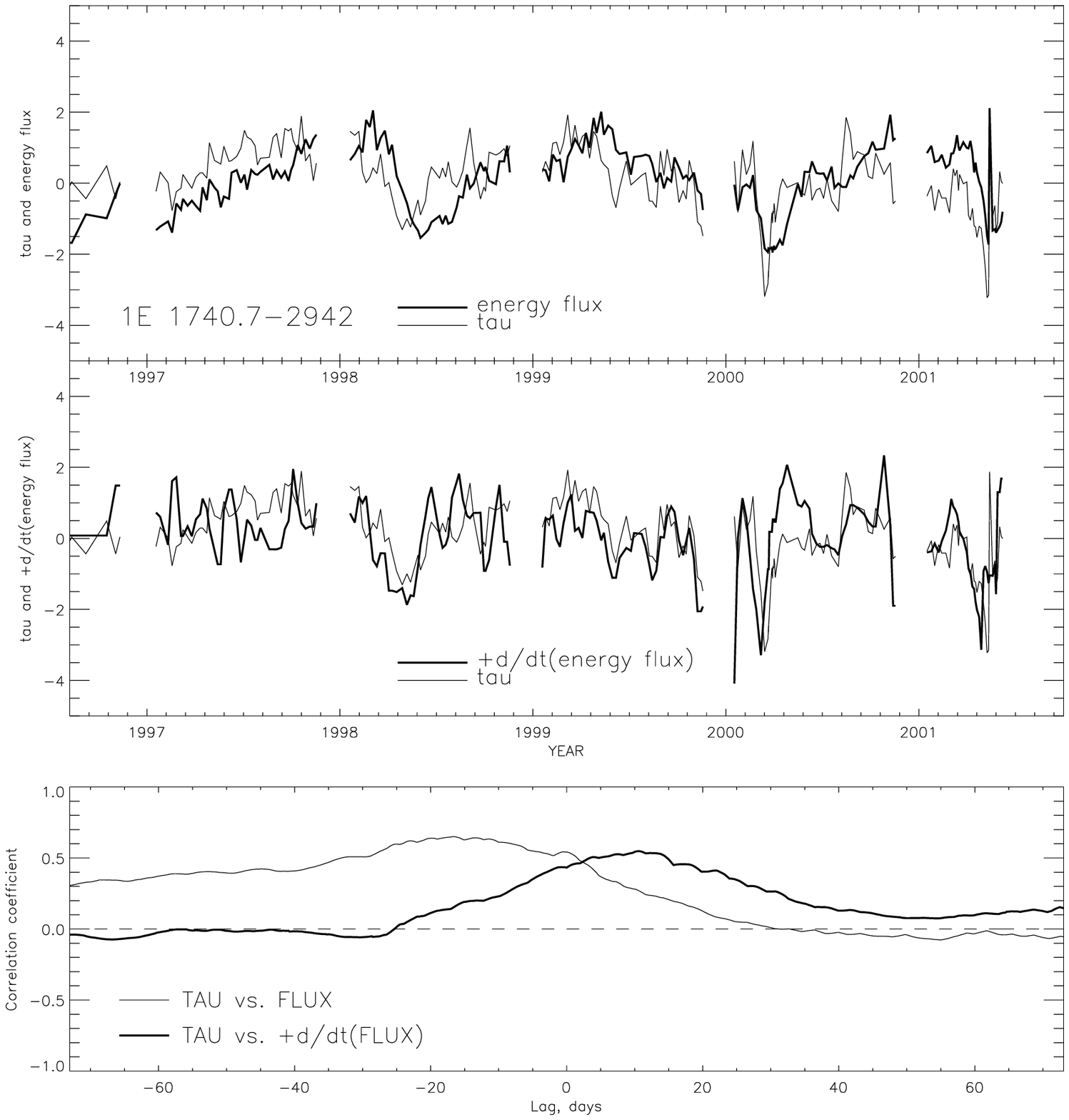}
\caption{Same as Figure 14 but with energy flux (2-100~keV) in place
of photon flux.}
\end{figure}

\begin{figure}
\plotone{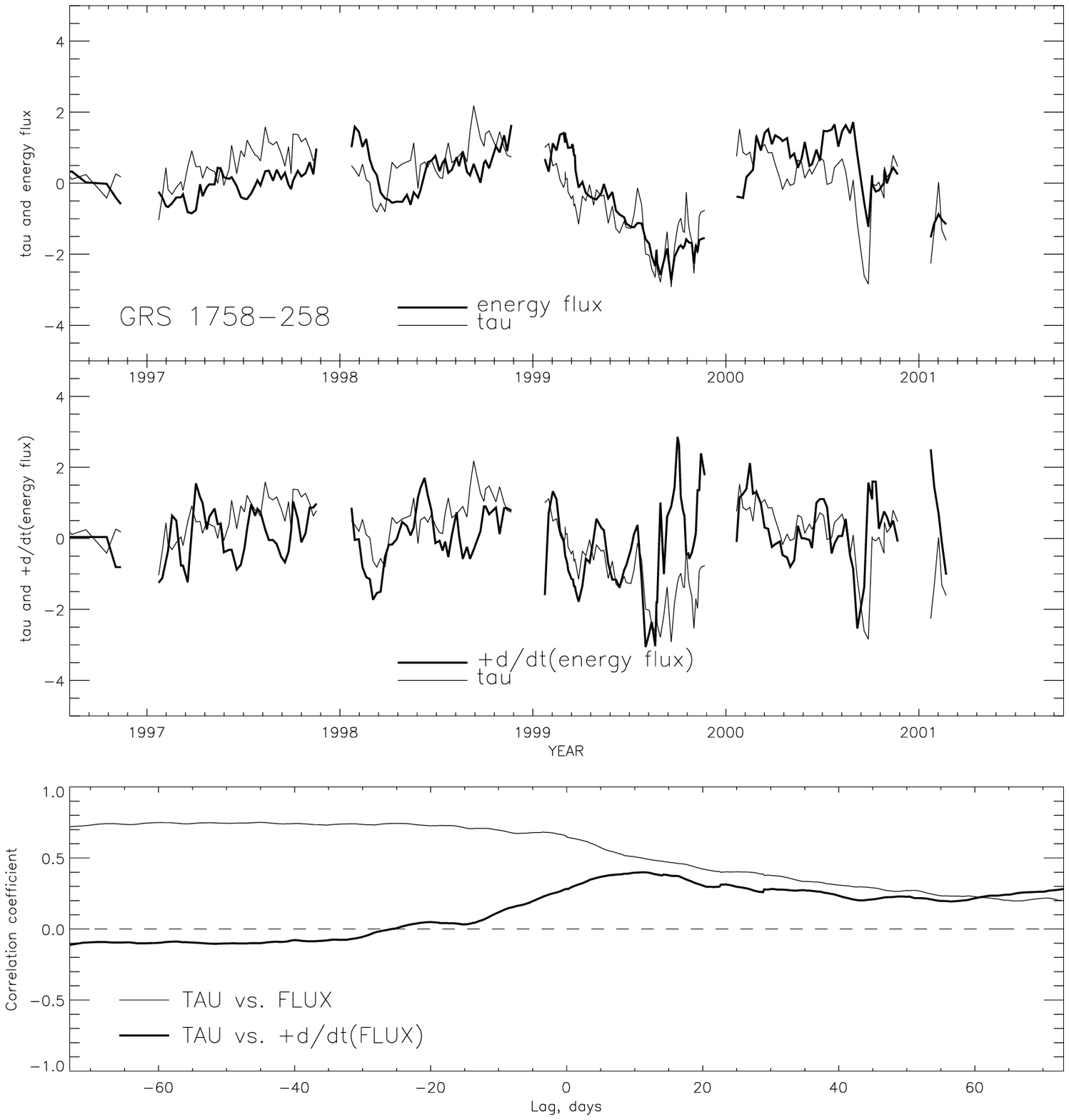}
\caption{Same as Figure 15 but with energy flux (2-100~keV) in place
of photon flux.}
\end{figure}

In unsaturated Comptonization, the PLI \citep{Sh76} is
\begin{equation}
\gamma \; = \; \left( \frac{9}{4} + \frac{4}{y(1+f)} \right) ^{1/2} \; - \; \frac{1}{2} 
\end{equation}
where the spectrum in photons cm$^{-2}$ s$^{-1}$ keV$^{-1} \sim E^{-\gamma}$.
The Comptonization parameter 
\begin{equation}
y \; = \; \frac{4kT_e}{m_e c^2} \rm{Max}(\tau,\tau^2)
\end{equation}
and $f$ is a function only of the halo electron temperature $T_e$:
\begin{equation}
f \; = \; 2.5\theta + 1.875\theta^{2}(1-\theta), \; \; \; \theta \equiv
k T_e / m_e c^2
\end{equation}
which we are taking as constant.  Then, at $k T_e = $100~keV,
\begin{equation}
\tau \; = \; 3.3 \left[ (\gamma + \frac{1}{2})^2 - \frac{9}{4}\right]^{-1}
\; \; \; \; \rm{or}  \; \; \; \; 
3.3^{1/2} \left[ (\gamma + \frac{1}{2})^2 - \frac{9}{4}\right]^{-1/2},
\end{equation}
whichever is less.  With this assumption, $\tau$ for \onee\ and
\grs\ remains close to 1, varying only from 0.44 to 1.39.
Since $\tau$ increases monotonically as 
$\gamma$ decreases, its correlations with photon flux, energy flux, and
their derivatives are of the opposite sign and similar magnitude
(Figures 13-17 and Table 1).

For an accreting halo flow with an unchanging geometry and constant
infall speed, $\tau$ is proportional to the accretion rate in this
flow, $\mdoth$.  Thus, if the geometry, infall speed, and temperature
of the halo are all constant, the PLI (via $\tau$) is a surrogate for
$\mdoth$.  If we could additionally postulate that the photon flux $F$
is a surrogate for the flow in the thin disk, $\mdotd$, and that the
hot halo is an independent flow, then Figures 13-15 could be
interpreted in simple and compelling ways.  For \onee\ and \grs, where
we have a good correlation in the second panel, we would have
\begin{equation}
\tau(t+\delta) - \overline{\tau} \;  \sim  \;  \frac{d}{dt}F(t) \; \; \; \rightarrow  \; \; \;  \mdoth(t+\delta) - \overline{\mdoth}\;  \sim  \; \frac{d}{dt} \mdotd(t),
\end{equation}
where $\delta < 0$ is the lag shown in Table 1.
The mean value is subtracted from $\tau$ because of the
offset to zero average imposed on all the curves in these figures.  This
offset has little effect on $dF/dt$, since
it already averages nearly to zero, there being no major long-term
trends in the count rates.

Equation 6 suggests that the flow in the thin disk at any time is an
integration of the halo flow in the recent past.  This would be
expected if there were two independent accretion flows: a thin disk
(with its slow time constant) and the halo (nearly in free fall), and
if they were supplied with proportional amounts of matter at large
radii at any given time.  The possibility of complicated variations of
PLI and luminosity as a result of independent disk and halo flows was
predicted by \citet{Ch95}.  Integration and delay are both natural
consequences of viscous propagation in a thin disk \citep{Fr92}.

Under these assumptions (constant halo temperature and
$F$ proportional to disk accretion rate), the observed
behavior of \cyg, with a strong inverse
correlation in the top panel of Figure 13, could result from
\begin{equation}
\mdoth(t) + \mdotd(t) \; \sim \; \rm{constant.}
\end{equation}
In other words, the state transitions could be due to the mass input
switching back and forth between going mostly into the halo flow
and mostly into the disk.  This is an appealing picture of \cyg\ for two
reasons.  First, the bolometric luminosity is, in fact,
very nearly constant during state transitions of \cyg\
\citep{Zh97}, which is not the case for other black-hole candidates.
Second, models which attribute the state changes to a
change in a single accretion rate \citep[e.g.,][]{Es98,Ja00} require
the average accretion rate to be in the narrow range of values
near where the hard-to-soft transition naturally occurs.  If
the state transitions are instead due to the swapping of the
accretion flow between two modes, they could occur over a
wide range of total accretion rates.

The principal problem with these simple and compelling pictures is the
use of $F$ as a surrogate for the thin disk flow.  If the
thin disk were a point source of soft photons at the center of a
large, optically thick Comptonizing cloud,
then every photon observed in the power law would have its origin in
the thin disk, and (nearly) every photon emitted by the thin disk
would end up in the power law.  Such a halo has been postulated by
\citet{Hu99} to explain the frequency dependence of lags between soft
and hard photons at sub-second timescales in black-hole candidates.
If the thin disk were a blackbody or a set of concentric blackbodies
at different temperatures, then $F$ (first from the thin
disk and, after Comptonization, in the power law) would go as
$\mdotd^{4/3}$, not very different from $\mdotd$.
For a halo which is close to the disk and tenuous, however, as is sometimes
the case in recent models, $\tau << 1$, most disk photons are not
Comptonized, and $F$ in the power law can be at
least as strong a function of $\mdoth$ as $\mdotd$.  A thin halo near
the disk also re-heats the disk with some of the Comptonized flux,
further complicating the issue.  That influence would be lessened if
the halo flow were geometrically large, so that most of the Comptonized flux
does not intersect the thin disk, or if it had relativistic bulk
motion away from the disk, so that the Comptonized photons moved
preferentially outward \citep{Be99}.

Finally, viscous action in the disk should not only delay but also
smooth out variations in the input to the accretion disk.  If the
$F$ were a good surrogate for the disk accretion flow, it
would therefore have less overall variability than the halo flow
(represented for the sake of this discussion by $\tau$).  This is not
the case, however.  For $\tau \lesssim 1$, the first exponent in
equation 5 is applicable.  In this case, the rms variability in $\tau$
and in $F$ are remarkably similar.  The rms variabilities
in $F$ and $\tau$ are 19\% and 18\% respectively in \onee\ and
22\% and 21\% in \grs.  This can't be seen in Figures 14-15 (top
panels) because of the scaling used, which automatically sets the rms
to 1, but the result is that for $\tau \lesssim 1$ the upper panels in
these figures would look the same even without the re-scaling.  For
$\tau > 1$, the second exponent in equation 5 applies and the problem
is aggravated: the rms variabilites in $\tau$ are only 9.4\% in \onee\
and 11\% in \grs, where we would expect them to be higher than the
variabilities in $F$.

On the other hand, the event of 2001
May 14 in \onee\ (Figure 12) involves both the PLI and energy flux
undergoing a large change while the photon flux remains unchanged.  This
shows that $F$ can sometimes have the slowest response
to sudden changes, an expected characteristic of the inner disk accretion
rate.

Our purpose, however, is not to argue that a very large, optically
thick halo with a uniform temperature actually exists.  Rather, it is
to stimulate more realistic theoretical work, and to point out that
any model of accretion in black-hole binaries must have at least the
potential to explain the behavior epitomized by Figure 8.  We have
used a model of two independent flows and the postulated
equivalence in equation 6 as a starting
point.  In a realistic two-flow model, the observables
$F$ and $\gamma$ would both be functions of both $\mdotd$ and $\mdoth$.
If the two-flow idea is valid, making these relations more physical
would result in maintaining the time lag between $\mdoth$ and $\mdotd$
while making the rms variability of the former greater than the
latter.

The disk and halo flows need not have an independent origin as they do
in \citet{Ch95}.  For example, \citet{Wa96} discussed how an initially
Keplerian corona, produced by evaporation from the disk, could lose
angular momentum and accrete rapidly due to radiation drag from a
central luminous source (i.e. a neutron star).  Greater luminosity
from the outer Keplerian thin disk would stabilize against this
effect.  For a black-hole system, we take the central radiation source
as being the innermost region of the disk itself.  If the mass input
to the thin disk dropped suddenly, the inner disk radiation would at
first dominate over the outer disk radiation since the drop would not
yet have propagated inward.  Thus the corona would accrete more
quickly than normal.  If the corona is produced at a constant rate by
evaporation, the result would be a thinner corona and a softer
power-law.  Once the drop in accretion rate had propagated to the
innermost regions, the former disk/corona ratio would re-establish
itself and the spectrum would harden again at a lower luminosity.
This reproduces the qualitative behavior in Figures 8 and 9: a
softening precedes a drop in flux.  What this picture has in common
with \citet{Ch95} is the viscous delay in the thin disk as the cause
of the complication in the observed behavior.

\section{Summary}

We have found that black-hole candidates \grs\ and \onee\ display
long-term behavior very different from the well-known variation
of \cyg.  The primary result, that the PLI varies as the negative derivative
of the power-law flux instead of directly with it, is at least suggestive
of a model with both slow (disk) and fast (halo) accretion flows.

Our observations come at a time when there is mounting evidence
that even the outer parts of accretion disks contain a vertically
extended flow as well as a standard thin disk.  \citet{Co01}
observed the low-mass x-ray binary EXO~0748--67 with the Reflection
Grating Spectrometer on the \it XMM-Newton \rm observatory.  Although
this is an eclipsing binary, they found very bright x-ray emission
lines which do not vary with the variation of the continuum from the
central source.  From this lack of variability and the details of the
absorption and emission in the spectrum, they concluded that there
must be a significant flow at large radii with a height much greater
than that predicted by hydrostatic equilibrium of a thin disk.  
\citet{Sm00} observed another dipping
binary, X1624--490, with \it RXTE\rm.  They found that the
Comptonization region which produces the hard x-ray continuum is never
completely occulted during dips, while the blackbody from the central
source is.  From the broad profile of the dips, they deduced that the
hard-x-ray-emitting region was $5 \times 10^{10}$ cm in radius or
larger, within a factor of two of the full presumed size of the thin
accretion disk based on the 21-hour orbit of the system.

\citet{Ka98} described certain correlations in the
atoll sources (neutron-star binaries) 4U~0614+091 and 4U~1608--52 that
might also be due to independent disk and halo flows.
In those sources, the frequency of the kilohertz quasi-periodic
oscillations (QPOs) correlates well with the power-law index
\citep{Ka98} and, for 4U~0614+091, with the disk blackbody
flux \citep{Fo97}.  Neither these sources, nor the kilohertz QPO
sources in general, show a long-term correlation between QPO frequency
and luminosity.  \citet{Ka98} suggested that the thin-disk
accretion rate may control the blackbody luminosity, the QPO frequency,
and (via Compton cooling of the halo) the power-law index, while
the overall luminosity (dominated by the power-law) is controlled
by the halo accretion rate.

Our secondary conclusion is that there can be two causes of spectral
softening in black-hole binaries: the well-modeled static soft state,
in which the accretion rate becomes high enough to force the innermost
part of the halo into the thin disk configuration, and a
newly-discovered dynamical soft state, occurring whenever the photon
flux is dropping.  In the context of a picture of two independent
accretion flows, the dynamical soft state would occur when the inner
thin disk has yet to respond to a drop in accretion which has already
depleted the halo.

Because persistent black hole candidates are rare, there are only a
few more immediately available tests of the ideas developed here.
LMC~X--3 is one source we have not yet examined.  It is too faint for
the \it RXTE \rm ASM data to be useful spectrally, and the PCA
monitoring observations so far, although they have revealed that the
source does undergo soft/hard transitions \citep{Wi01}, have been too
few to provide evidence for or against viscous delay.  With a main
sequence companion of only modestly high mass (B3 V), it is probably a
disk accretor, and may therefore be a good candidate for delays.  \gx,
which we expect to show delays, may provide more conclusive data than
the hint visible in Figure 4.  A new PCA monitoring campaign for this
source began in 2001 March.  Finally, the \it XMM-Newton \rm x-ray
telescope has sufficient sensitivity to measure spectra of sources as
faint as \onee\ and \grs\ in M31 with a reasonable exposure time ($<$
1 dy).  A monitoring campaign of the M31 bulge region could reveal a
number of new sources with behavior similar to \onee\ and \grs.  If
this behavior is indeed due to viscous delay in the disk, this
campaign would produce a census of high-mass versus low-mass binaries
using viscous delay as a distinguishing diagnostic.

In future work we will consider the connection between the behavior
exhibited by \onee\ and \grs\ and that observed in the x-ray novae or
soft x-ray transients.  The transient systems, exemplified by Nova
Muscae 1991 (GS~1124--683), tend to brighten in their hard state,
change to the soft state near their peak, and decay in the soft state,
returning to the hard state only at much lower luminosities
\citep{Mi95}.  \gx, much more variable than \onee\ and \grs\ but more
persistent than the transients, may serve to clarify the connection.

The authors thank two anonymous referees for many useful suggestions
and corrections.  This work was funded in part by NASA grant
NAG5-7265.

\end{document}